\documentclass[aas_macros]{aa}
\usepackage{times}
\usepackage{graphicx}
\usepackage{subfigure}
\usepackage{amsmath}
\usepackage{color}
\usepackage{xcolor}
\usepackage{multirow}
\usepackage{verbatim}
\usepackage{multicol}
\usepackage{float}
\usepackage{tabularx}
\usepackage[normalem]{ulem}
\bibpunct{(}{)}{;}{a}{}{,} % to follow the A&A style
\usepackage[colorlinks=true,linkcolor=red,citecolor=blue, breaklinks=true]{hyperref}
\usepackage{natbib,twoopt}
\usepackage[capitalise,noabbrev]{cleveref}
\graphicspath{{Figs/}}

\makeatletter
\newcommandtwoopt{\citeads}[3][][]{\href{http://adsabs.harvard.edu/abs/#3}%
{\def\hyper@linkstart##1##2{}%
\let\hyper@linkend\@empty\citealp[#1][#2]{#3}}}
\newcommandtwoopt{\citepads}[3][][]{\href{http://adsabs.harvard.edu/abs/#3}%
{\def\hyper@linkstart##1##2{}%
\let\hyper@linkend\@empty\citep[#1][#2]{#3}}}
\newcommandtwoopt{\citetads}[3][][]{\href{http://adsabs.harvard.edu/abs/#3}%
{\def\hyper@linkstart##1##2{}%
\let\hyper@linkend\@empty\citet[#1][#2]{#3}}}
\newcommandtwoopt{\citeyearads}[3][][]%
{\href{http://adsabs.harvard.edu/abs/#3}
{\def\hyper@linkstart##1##2{}%
\let\hyper@linkend\@empty\citeyear[#1][#2]{#3}}}
\makeatother

\providecommand{\sorthelp}[1]{}
\newbox\tablebox    \newdimen\tablewidth
\def\leaderfil{\leaders\hbox to 5pt{\hss.\hss}\hfil}

\def\tablenote#1 #2\par{\begingroup \parindent=0.8em
    \abovedisplayshortskip=0pt\belowdisplayshortskip=0pt
    \noindent
    $$\hss\vbox{\hsize\tablewidth \hangindent=\parindent \hangafter=1 \noindent
    \hbox to \parindent{$^#1$\hss}\strut#2\strut\par}\hss$$
    \endgroup}

\def\deg{\ifmmode^\circ\else$^\circ$\fi}
\def\pdeg{\ifmmode $\setbox0=\hbox{$^{\circ}$}\rlap{\hskip.11\wd0 .}$^{\circ}
          \else \setbox0=\hbox{$^{\circ}$}\rlap{\hskip.11\wd0 .}$^{\circ}$\fi}
\def\arcm{\ifmmode {^{\scriptstyle\prime}}
          \else $^{\scriptstyle\prime}$\fi}

\DeclareMathAlphabet{\mathsc}{OT1}{cmr}{m}{sc}
\def\testbx{bx}%
\DeclareRobustCommand{\ion}[2]{%
\relax\ifmmode
\ifx\testbx\f@series
{\mathbf{#1\,\mathsc{#2}}}\else
{\mathrm{#1\,\mathsc{#2}}}\fi
\else\textup{#1\,{\mdseries\textsc{#2}}}%
\fi}

\newcommand{\pasiphae}{\textit{PASIPHAE\ }}
\def\deg{\ifmmode^\circ\else$^\circ$\fi}
\def\arcs{$^{\scriptstyle\prime\prime}$}
\newcommand{\SNR}{\ensuremath{\mathrm{S/N}}\xspace}
\newcommand{\texp}{\ensuremath{t_{\mathrm{exp}}}\xspace}

\begin{document}

\title{WALOP-South: a four camera one shot imaging polarimeter for the  \pasiphae survey. Paper \textsc{III} -- PSF modelling}
\author{Indrajit Paul\thanks{Corresponding author: indrajit.paul@niser.ac.in} \inst{1,2}, Kishan Deka\thanks{Corresponding author:  kishan.deka@ncbj.gov.pl} \inst{3}, Tuhin Ghosh \inst{1,2}, Siddharth Maharana \inst{4,12}, A.N. Ramaprakash \inst{4,6,5}, Ramya M. Anche \inst{4,7},  Artem Basyrov \inst{8}, Dmitry Blinov \inst{6,9}, 
Hans Kritian  Eriksen \inst{8}, Eirik Gjerl\o w \inst{8}, John A. Kypriotakis \inst{6,9}, Sebastian Kiehlmann \inst{6,9}, Ioannis Liodakis \inst{6}, Nikolaos Mandarakas \inst{6,9}, Georgia V. Panopoulou \inst{10}, Vasiliki Pavlidou \inst{6,9}, Timothy J. Pearson \inst{5}, Vincent Pelgrims \inst{11}, Stephen B. Potter \inst{12,13}, Anthony C. S. Readhead \inst{5}, Raphael Skalidis \inst{14}, Trygve Leithe Svalheim \inst{8}, Konstantinos Tassis \inst{6,9}, Namita Uppal \inst{6}, Ingunn K. Wehus \inst{8}}
\institute{School for Physical Sciences, National Institute of Science Education and Research, HBNI Jatni-752050, India
\and
Homi Bhabha National Institute, Training School Complex, Anushakti Nagar, Mumbai 400094, India
\and
Astrophysics division, National Center for Nuclear Research, Pasteura 7, 02093, Warsaw, Poland 
\and
Inter-University Centre for Astronomy and Astrophysics, Post bag 4, Ganeshkhind, Pune, 411007, India 
\and
Cahill Center for Astronomy and Astrophysics, California Institute of Technology, Pasadena, CA, 91125, USA 
\and
Institute of Astrophysics, Foundation for Research and Technology-Hellas, Voutes, 70013 Heraklion, Greece
\and
Steward Observatory, University of Arizona, Tucson, Arizona, 85721, USA
\and
Institute of Theoretical Astrophysics, University of Oslo, P.O. Box 1029 Blindern, NO-0315 Oslo, Norway
\and
Department of Physics, University of Crete, Voutes, 70013 Heraklion, Greece
\and
Department of Space, Earth and Environment, Chalmers University of Technology, 412 93, G\"{o}teborg, Sweden
\and
Universit{\'e} Libre de Bruxelles, Science Faculty CP230, B-1050 Brussels, Belgium 
\and
South African Astronomical Observatory, PO Box 9, Observatory, 7935, Cape Town, South Africa
\and
Department of Physics, University of Johannesburg, PO Box 524, Auckland Park 2006, South Africa 
\and
Owens Valley Radio Observatory, California Institute of Technology, MC 249-17, Pasadena, CA 91125, USA}

\abstract{The two WALOP instruments, built for the \pasiphae survey, will measure the linear polarization of large numbers of stars in the Galactic polar regions in the SDSS-$r$ band. They are designed with a wide field-of-view, enabling measurement of the Stokes parameters $I$, $q$, and $u$ for multiple stars simultaneously within a $35\arcmin \times 35\arcmin$ region of sky for WALOP-South and $30\arcmin \times 30\arcmin$ region for WALOP-North. In this paper, we present a polar shapelet-based PSF photometry framework for well-sampled stellar point sources applicable to WALOP-type wide-field polarimeters. Polar shapelets are a set of orthogonal basis functions, constructed from Gauss-Hermite or Gauss-Laguerre polynomials, that are well-suited to modelling localized PSF in a compact and efficient way. We developed an efficient PSF modelling method that uses polar shapelets as basis functions to reconstruct the spatial variation of the PSF shape across the CCD using \textsc{Zemax}-simulated images of one of the WALOP instruments, and show that a limited number of shapelet coefficients are sufficient to capture this variation consistently across different CCD locations. To simulate realistic star images, we introduce random sub-pixel shifts in the star centroids in the \textsc{Zemax}-simulated images, and account for this using a two-step iterative method that alternately estimates the PSF model and the sub-pixel centroid shift. Applying PSF photometry to the target faint stars, we demonstrate that the photometric accuracy of approximately 0.15\% is achievable, and that the reconstructed PSF model can be incorporated into the photometry across different seeing conditions, meeting the polarimetric science requirements of the \pasiphae survey.}

\titlerunning{PSF modelling for WALOP-South instrument} 
\authorrunning{Paul et al.} 
\maketitle

\section{Introduction}   \label{sec:sect1}

Linear polarization measurements of stars offer a unique probe of the magnetic field structure in the intervening dust clouds of the Milky Way. Upcoming high-sensitivity polarimetric surveys pursuing such scientific objectives require high photometric accuracy. Aperture photometry is the most widely used technique for measuring the total photon count of relatively bright, isolated stars. In this method, all pixels are weighted equally, and the uncertainty per pixel is taken as the square root of the photon count under the assumption of Poisson-distributed noise. For distorted or asymmetric star profiles, however, aperture photometry fails to recover the true photon count accurately, a limitation that is further exacerbated for faint sources \citep{Handler:2003, Becker:2007}. In such cases, the photon count is instead estimated using the Point Spread Function (PSF), a two-dimensional (2D) profile describing the spatial distribution of photons from a point source on the detector \citep{Anderson:2000, Mighell:2005}. The PSF is used to assign a weight to each pixel when computing the total photon count, thereby reducing the photometric uncertainty and enabling the desired accuracy to be achieved. For overlapping stars in crowded fields, PSF photometry outperforms aperture photometry \citep{Stetson:1987}.

PSF shapes captured on the detector are determined by atmospheric conditions, the optics of the telescope and the instrument, pixel sensitivity, and the pixel response function of the detector. In the literature, we find a detailed discussion about the \textit{instrumental} PSF caused by the optics of an instrument and the \textit{effective} PSF captured by the detector. Here, we consider only the \textit{effective} PSF and we refer to it simply as PSF in the scope of this paper. The first step is to model the PSF shape at the exact location where the star image has formed. Near the axis of the detector, the PSF can often be approximated by a 2D Gaussian function. But in the case of wide-field observations the PSF shapes get deformed as we move away from the optical axis due to aberrations, vignetting etc. So, we require a robust method that can model both the PSF and its variation across the field-of-view. The second step is to use the PSF model at the CCD position where the star image is formed as a weight function to count the total number of photons.

A method for PSF modelling and PSF photometry is demonstrated in the context of the Polar-Areas Stellar Imaging in Polarization High-Accuracy Experiment (\textit{PASIPHAE}), an upcoming optopolarimetric survey \citep{Tassis:2018}. Planned by an international collaboration, the \pasiphae survey aims to measure the linear polarization of millions of stars and to use these measurements to construct a three-dimensional tomographic map of the magnetic field threading dust clouds within the Milky Way. To produce a large-area optical polarization map of the sky, two wide-field, high-accuracy polarimeters, named Wide Area Linear Optical Polarimeters (WALOPs) are currently under development at the Inter-University Centre for Astronomy and Astrophysics, Pune, India. These two instruments, WALOP-South and WALOP-North, are the successors to the RoboPol polarimeter  \citep{Ramaprakash:2019}, which has been operating at the Skinakas Observatory, Crete, Greece since 2013. WALOP-South will be installed on the $1$\,m telescope at the Sutherland site of the South African Astronomical Observatory (SAAO), South Africa \citep{walop_s_spie_2020, Maharana:2021}, and WALOP-North on the $1.3$\,m telescope at the Skinakas Observatory, Crete, Greece \citep{Kypriotakis:2024}. The photometric analysis presented here is limited to the optical model of WALOP-South, though the methods are equally applicable to WALOP-North. With a field-of-view of $35^\prime \times 35^\prime$ for WALOP-South and $30^\prime \times 30^\prime$ for WALOP-North, each instrument produces four simultaneous images of every source, enabling the determination of the magnitude, linear fractional polarization ($p$), and polarization angle ($\theta$) in the SDSS-$r$ filter. The \pasiphae survey aims to achieve a target polarimetric accuracy of $\sigma_p = 0.1\%$ with the WALOP instruments. The calibration methodology for WALOP-South, based on simulated data and designed to achieve the target polarimetric accuracy, is presented in \citet{Maharana:2022}.

Characterising the PSF shape and its variation across the focal plane is extremely challenging. Analytic modelling of the PSF by fitting a large number of free parameters is not feasible owing to the substantial variation in PSF shapes across the wide field-of-view of the WALOP-South instrument. \citet{Stetson:1987} demonstrated a hybrid method combining an elliptical Gaussian fit with an interpolation scheme to capture the PSF residuals. However, a simple Gaussian fit is insufficient for the complex PSF shapes produced by the WALOP-South instrument. A more sophisticated approach, employing Karhunen--Lo\`{e}ve basis functions for PSF modelling, was developed for the Sloan Digital Sky Survey (SDSS) by \citet{2001ASPC..238..269L}. More recently, precise PSF modelling has been achieved by decomposing the PSF into a set of basis functions. \citet{2013A&A...551A.119P} adopted this approach using modified Zernike polynomials as basis functions. Gaussian-weighted Hermite polynomials (polar shapelets) have been employed to characterise galaxy morphologies in Hubble Space Telescope (HST) imaging \citep{Massey:2005}. In weak lensing studies, small distortions in galaxy images have been modelled using polar shapelets \citep{2003MNRAS.338...35R, 2003MNRAS.338...48R}. Polar shapelets have further been applied to the morphological classification of SDSS 
galaxies \citep{Kelly_2004}.

In this work, the polar shapelet formalism for PSF modelling is applied to develop a robust technique for accurate photon count measurements from distorted star images of the WALOP-South instrument. The polar shapelet basis is both complete and orthogonal, ensuring computational efficiency and accurate decomposition.  Unlike other PSF modelling approaches, a single set of fixed basis parameters is sufficient to represent the full range of PSF shapes across the detector, enabling the same basis functions to be applied to all PSF shapes within a single frame and significantly reducing the overall computational cost. The aim of this work is to determine the number of shapelet coefficients required to capture the full range of PSF shape variations and to achieve the target photometric accuracy.

This work specifically addresses two challenges in PSF modelling and PSF photometry: sub-pixel centroid shifts and variable atmospheric seeing. When the centroids of the stars used to construct the PSF deviate by even a fraction of a pixel, the modelling accuracy can degrade. This effect can be mitigated by explicitly accounting for the sub-pixel level centroid offsets of the star used for PSF modelling \citep{Lauer_1999, Barron_2007, Wu_2025}. For ground-based observations, atmospheric turbulence causes each star point source to be broadened into an approximately Gaussian profile whose full width at half maximum (FWHM) is determined by the prevailing seeing conditions. The observed star image on the detector is therefore well described by the convolution of the true optical PSF at the given detector position with a 2D Gaussian of appropriate FWHM. This complicates PSF modelling, as the effective PSF at a fixed detector position broadens with worsening seeing and thus varies between observations taken at different epochs. To address this, a robust method is presented in which the PSF is characterised once under optimal seeing conditions, and the properties of convolution are subsequently exploited to predict the effective PSF under variable or degraded seeing.

This paper is organised as follows. Section~\ref{sec:2} provides an overview of the WALOP-South instrument and its use in measuring starlight polarization. Section~\ref{sec:methodology} describes the core analytical framework, encompassing the polar shapelet formalism, the PSF modelling technique, and the application of PSF photometry under variable seeing conditions. Sect.~\ref{sec:simulations} details the preparation of simulated observations for WALOP-South using the \textsc{Zemax} optical design software. The results of the simulated PSFs for WALOP-South are presented in Sect.~\ref{sec:results}. Finally, Sect.~\ref{sec:conclusion} summarises our findings.

\section{Polarimetry with WALOP-South}\label{sec:2}

WALOP-South is a one-shot, four-channel linear optical polarimeter. A detailed description of the optical model of WALOP-South is presented in \cite{Maharana:2021}. The instrument comprises a collimator assembly that accepts the wide field-of-view beam from the telescope focal plane and generates a pupil image. The polarizer assembly then splits the pupil beam into four polarised beams (O1, O2, E1, and E2), directing them along the $\pm x$ and $\pm y$ directions. The O1 and O2 beams, directed along $+y$ and $-y$ respectively, correspond to polarization angles of $0\degr$ and $45\degr$, while the E1 and E2 beams, directed along $-x$ and $+x$ respectively, correspond to $90\degr$ and $135\degr$. The polarizer assembly consists of BK7 glass wedges, half-wave retarder plates, and Wollaston prisms, together constituting the polarization analyser unit of the instrument. Each beam is focused onto a dedicated $4096 \times 4096$ pixel detector by an individual camera assembly, enabling simultaneous imaging of the full field-of-view in all four polarization channels.

For each star in the field, an image is formed on each of the four detectors. Differential photometry across the four channels yields the normalised Stokes parameters\footnote{The normalised Stokes parameters are defined as $q = Q/I$ and $u = U/I$, where $I$ is the photon intensity and $Q$ and $U$ are the linear polarization Stokes parameters.} ($q$ and $u$) and their associated uncertainties,
\begin{align}
    q &= \frac{I_{0\degr} - I_{90\degr}}{I_{0\degr} + I_{90\degr}}, 
    \quad 
    &&\sigma_{q} = \sqrt{\frac{4\left(I_{90\degr}^{2}\,\sigma_{I_{0\degr}}^{2} 
    + I_{0\degr}^{2}\,\sigma_{I_{90\degr}}^{2}\right)}
    {\left(I_{0\degr}+I_{90\degr}\right)^{4}}} \, , \label{eq:2.1} \\
    u &= \frac{I_{45\degr} - I_{135\degr}}{I_{45\degr} + I_{135\degr}},
    \quad
    &&\sigma_{u} = \sqrt{\frac{4\left(I_{45\degr}^{2}\,\sigma_{I_{135\degr}}^{2} 
    + I_{135\degr}^{2}\,\sigma_{I_{45\degr}}^{2}\right)}
    {\left(I_{45\degr}+I_{135\degr}\right)^{4}}} \, , \label{eq:2.2}
\end{align}
where $I_{0\degr}$, $I_{90\degr}$, $I_{45\degr}$, and $I_{135\degr}$ are the star photon counts measured in the four polarised channels, and $\sigma_{I_{0\degr}}$, $\sigma_{I_{90\degr}}$, $\sigma_{I_{45\degr}}$, and 
$\sigma_{I_{135\degr}}$ are the corresponding photometric uncertainties \citep{Ramaprakash:2019}. From the measured Stokes parameters $q$ and $u$, the degree of linear polarization $p$ and the polarization angle $\theta$ can be derived as
\begin{equation}
    p = \sqrt{q^2 + u^2}\,, \qquad \theta = \frac{1}{2}\arctan\left(\frac{u}{q}\right) \, .
\end{equation}

To achieve a polarimetric accuracy of $\sigma_p = 0.1\%$, the required photometric accuracy is $\sigma_I / I \sim 0.15\%$, where $I$ is the total photon count measured in a single detector channel. This estimate is obtained by considering the case in which the $q$ measurement attains its maximum value, corresponding to a polarization degree of $p = 0.01$ with $u = 0$. Under this configuration, the total count ratio $I_{90\degr}/I_{0\degr} = 0.98$, and Eq.~(\ref{eq:2.1}) yields the relation between photometric and polarimetric accuracy. An analogous relation is obtained by considering the case in which $u$ attains 
its maximum value with $q = 0$, using Eq.~(\ref{eq:2.2}).

\section{Methodology} \label{sec:methodology}

For a given star, the observed image represents the distribution of photons across a 2D pixel grid. Adopting a polar coordinate system $(r, \theta)$, the image $D$ of a single star can be expressed as
\begin{equation}
D(r,\theta) = I_\star\, M(r,\theta;\, x^c, y^c) + I_b(r,\theta) \, , \label{eq:3.8}
\end{equation}
where $I_\star$ is the total photon count of the star (including Poisson noise), $M$ is the normalised PSF, and $(x^c, y^c)$ denotes the sub-pixel displacement of the PSF centroid from the image centre, with $(x^c, y^c) = (0, 0)$ corresponding to a perfectly centred source. The term $I_b$ accounts for the sky background contribution and its associated Poisson noise.

The mean sky background per pixel, $\langle I_b \rangle$, is estimated as the $\sigma$-clipped median of the photon counts within an annulus centred on the star but sufficiently distant to avoid contamination from its wings \citep{Stetson:1987, SExtractor+1996}. This $\sigma$-clipping procedure ensures robustness against outliers and defective pixels.

Following background subtraction to obtain the processed image $\tilde{D}$ ($= D - \langle I_b \rangle$), 
we determine the total photon count $I_\star$ and the centroid coordinates 
$(x^c, y^c)$ by minimising the loss function $S$, given by
\begin{equation}
    S = \sum_{r,\theta} \frac{\left[\tilde{D}(r,\theta) - 
    I_\star\, M(r,\theta;\, x^c, y^c)\right]^2}{\sigma^2(r,\theta)} 
    \, , \label{eq:3.10}
\end{equation}
where $\sigma^2(r,\theta)$ is the pixel-level variance. Although photon counts strictly follow a Poisson distribution, the high $\SNR$ of our observations justifies a Gaussian approximation, as the Poisson distribution converges to a Gaussian in the high-count limit by the central limit theorem.

This estimation requires \textit{a priori} knowledge of the normalised PSF; therefore, our primary goal is to characterise the PSF accurately. This is particularly challenging for WALOP-South owing to its wide field-of-view. In certain regions of the focal plane, the light beam deviates significantly from the optical axis, causing star profiles to deviate from a standard 2D Gaussian.

To capture these spatial variations, we model the PSF using polar shapelets. Normalised images of multiple bright stars with high $\SNR$ are used to reconstruct a stable and accurate model PSF \citep{heasley1999point}. This model PSF is subsequently applied to perform PSF photometry on target stars having low $\SNR$, ensuring reliable flux extraction even in the presence of significant distortion and noise.

\subsection{Polar shapelets} \label{sec:sect3.1}

The polar shapelets formalism decomposes any localized object into a series of orthogonal basis functions with suitable weights. The polar shapelet basis functions $\chi _{n,m}(r, \theta;\beta)$ are orthonormal and complete in the pixel space \citep{Massey:2005}. They are parameterized  by  two  integers, $n$ and $m$, and they are given by
\begin{multline}
        \chi_{n,m}(r,\theta;\beta)= \frac{(-1)^{\frac{n-|m|}{2}}}{\beta ^{|m|+1}}  \Bigg[\frac{\Big(\frac{n-|m|}{2}\Big]!}{\pi\Big(\frac{n+|m|}{2}\Big)! }\Bigg]^{\frac{1}{2}}  r^{|m|}L^{|m|}_{\frac{n-|m|}{2}} \Bigg(\frac{r^2}{\beta^2}\Bigg)\\ 
        \times  e^{-r^2/2\beta^2}e^ {-i m\theta} \ ,  \label{eq:3.1}
\end{multline}
where $(r, \theta)$ represents the polar coordinate of the pixel, $\beta$ is the scale factor, and $L_l^{|m|}$ are associated Laguerre polynomials. We chose the coordinate system such that the origin is at the centre of the image. $\chi_{n,m}(r,\theta;\beta)$ can be obtained at any location with centre not in the origin (say, $x^c, y^c$) using the coordinate transformation $r=\sqrt{(x-x^c)^2+(y-y^c)^2}$ and $\theta = \arctan((y-y^c)/(x-x^c))$. By construction, the basis functions $\chi_{n,m}$ satisfy the following symmetry conditions, 
\begin{equation}  
\chi_{n,-m}(r, \theta;\beta) = \chi^*_{n,m}(r, \theta;\beta) = \chi_{n,m}(r, -\theta;\beta)  \ . \label{eq:3.2}
\end{equation}
Polar shapelets form a complete basis set in the complex plane and satisfies the orthogonality condition,
\begin{equation}
 \int \int \chi _{n,m}^*(r, \theta;\beta) \chi _{u,v}(r, \theta;\beta) r dr d\theta= \delta_{nm}\delta_{uv} \ , \label{eq:3.3}
\end{equation}
where $\delta$ is the Kronecker delta and $*$ denotes complex conjugation. The basis function $\chi_{n,m}$ is only defined when $(n-|m|)/2$ is a non-negative integer. That is why the real part of $\chi_{n,0}$ vanishes when $n$ is odd. For $m=0$, we define $\chi_{n,0}$ as \textit{central} terms and they show azimuthal symmetry. 
The polar shapelet basis functions (real and imaginary components) for different values of $n$ and $m$, truncated at $n_{\rm max} = 5$, are shown in Appendix~\ref{appendix:A}; for higher orders, only the real component of the central basis functions is shown.

\subsection{PSF modelling}\label{sec:sect3.2}

A localised 2D  PSF, defined as $d(r, \theta) = \tilde{D}(r, \theta) / \sum_{r, \theta} \tilde{D}(r, \theta)$, can be decomposed in terms of the polar shapelet basis \citep{Massey:2005}. The basis coefficients $b_{n,m}$ are obtained by projecting the PSF onto the shapelet basis functions $\chi_{n,m}(r, \theta; \beta)$:
\begin{equation}
    b_{n,m} = \iint d(r, \theta)\, \chi^*_{n,m}(r, \theta; \beta)\, 
    r\, dr\, d\theta \, , \label{eq:3.4}
\end{equation}
with
\begin{equation}
d(r, \theta) = \sum_{n=0}^{\infty} \sum_{m=-n}^{n} b_{n,m}\, \chi_{n,m}(r, \theta; \beta) \, , \label{eq:3.5}
\end{equation}
where $n$ is a non-negative integer, $m$ is an integer ranging from $-n$ to $n$ in steps of two, and $r$ is the radial distance from the PSF centre. The integral in Eq.~(\ref{eq:3.4}) is approximated by a discrete sum over the pixels comprising the PSF image, with pixel area $s$:
\begin{equation}
b_{n,m} \approx s \sum_{r,\theta} d(r, \theta)\, \chi^*_{n,m}(r, \theta; \beta) \, . \label{eq:3.6}
\end{equation}

\begin{figure}
\centering
\includegraphics[width=\linewidth]{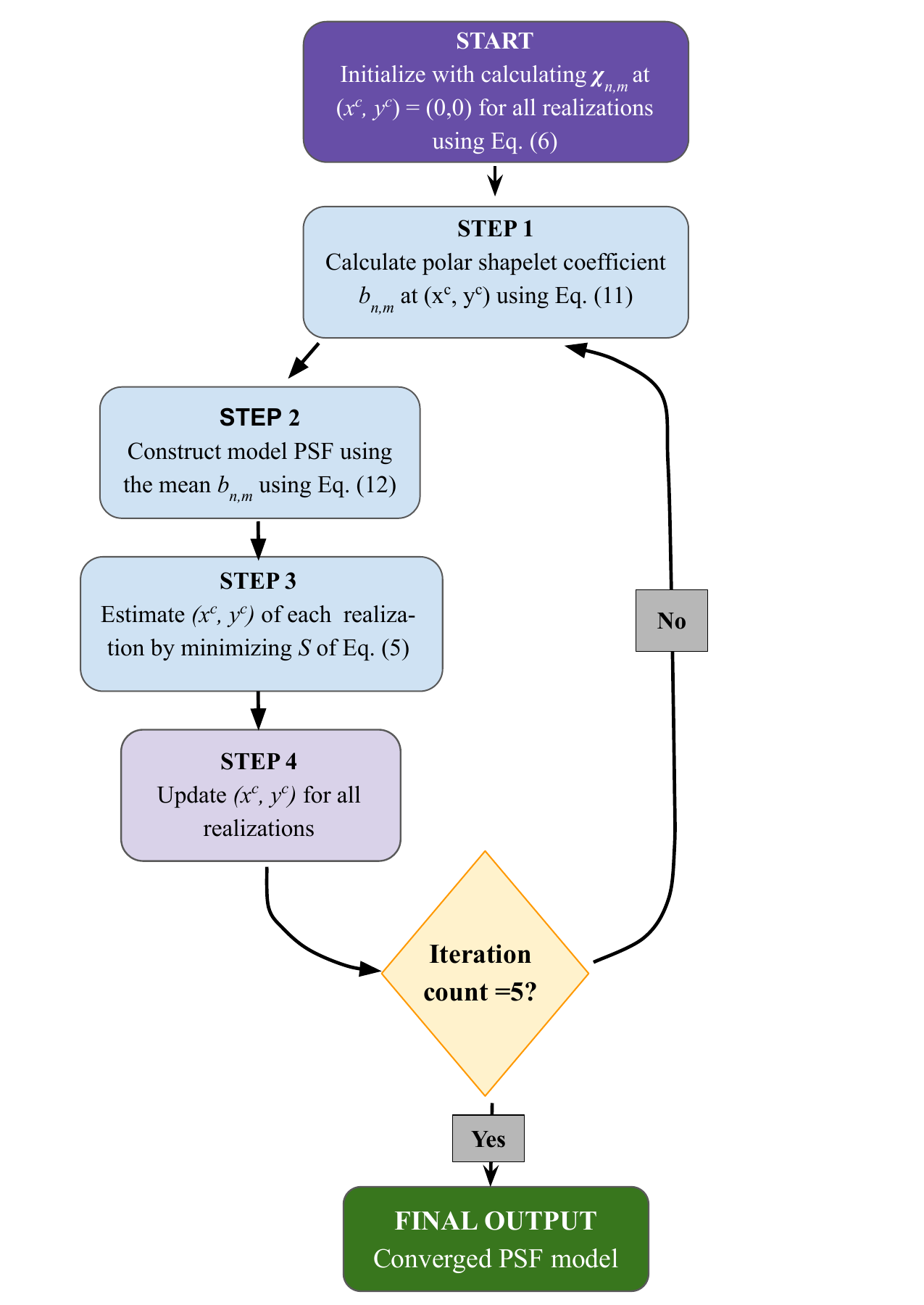}
\caption{Flowchart of the iterative self-consistent PSF modelling algorithm. The process alternates between (1) estimating the mean shapelet coefficients from multiple realisations and (2) refining the sub-pixel centroid positions by minimising the squared residuals against the current model. The cycle repeats for five iterations to ensure parameter convergence.}
\label{fig:flowchart}
\end{figure}

While the PSF of a single star is sufficient to compute the shapelet coefficients, using multiple realisations of the star PSF helps mitigate the effects of noise and spatial variations in the PSF shape across the CCD.  In practice, however, the true star centroid often lies at an arbitrary sub-pixel position. To account for this, a two-step iterative procedure is employed to determine the PSF model and the centroid position self-consistently, as illustrated in Fig.~\ref{fig:flowchart}.

In the first step, the polar shapelet basis $\chi_{n,m}$ is constructed with the same dimensions as the image using Eq.~(\ref{eq:3.1}), with the centroid initially fixed at the origin $(x^c, y^c) = (0, 0)$, for each realisation. By setting $s = 1$, the complex shapelet coefficients $b_{n,m}$ are obtained by computing the pixel-wise product of $d$ and $\chi^*_{n,m}$ and summing over all pixels in the image. The mean coefficients are then computed across all realisations, yielding a model PSF:
\begin{align}
M(r,\theta) &\approx \sum_{n=0}^{n_{\rm max}} \sum_{m=-n}^{n} 
    b_{n,m}\,\chi_{n,m} \nonumber \\
&= \sum_{n=0}^{n_{\rm max}} \sum_{m=-n}^{n}
    \left( b^{r}_{n,m}\,\chi^{r}_{n,m} - b^{i}_{n,m}\,\chi^{i}_{n,m} 
    \right) \, , \label{eq:3.13}
\end{align}
where the superscripts $r$ and $i$ denote the real and imaginary components, respectively.

In the second step, the model PSF is used to determine the sub-pixel centroid coordinates $(x^c, y^c)$ by minimising the loss function defined in Eq.~(\ref{eq:3.10}). For this positional refinement, the variance is set to unity, treating the minimisation as an ``\textit{ordinary least squares}'' problem. Adopting a uniform weight ensures that the steep gradients in the star core dominate the minimisation, which has been shown to yield a more stable and unbiased centroid estimate \citep{Stetson:1987, Mighell+1999}. To prevent flux variations from biasing the model, each image is normalised such that $I_\star = 1$. The updated centroids are subsequently fed back into the iterative loop to recompute the shapelet coefficients.

This cycle repeats until both the centroid positions and shapelet coefficients converge. In practice, convergence is typically reached within four to five iterations; a fixed limit of five iterations is therefore adopted for all PSF models. This strategy improves numerical stability and efficiency by avoiding the simultaneous minimisation of a high-dimensional parameter space. The minimisation of the non-linear loss function $S$ is performed using Powell's direction-set method \citep{Powell1964}, which is well-suited for discrete image data and Fourier-domain sub-pixel shifts and implemented in the SciPy\footnote{\href{https://scipy.org/}{https://scipy.org/}} package \citep{Virtanen2020}.

\subsection{Including variable seeing conditions in PSF modelling} \label{sec:sect3.3}

To model the PSF shapes in presence of varying seeing conditions, we used the properties of convolution instead of using polar shapelet modelling for every different observing conditions. First, we considered a median seeing condition when broadening effect on the image is optimal at the observation site. We model the PSF shape using observations at median seeing condition with the help of polar shapelets. Then we convolved this polar shapelet model with an appropriate Gaussian distribution to obtain the other broadened PSF shapes for worse seeing conditions. 

The mathematical framework of decoupling the seeing effect from broadened PSF shapes is simple. Consider $\tilde M(r,\theta)$ is the PSF model of a star image at median seeing. The polar shapelet model is defined by  Eq.~\eqref{eq:3.13}, where the centre is taken to be at origin. The PSF shape at median seeing, $\widetilde{M}(r,\theta)$ can be written as convolution of the true point source PSF and a Gaussian function of spread, $\sigma_1$. Similarly, the PSF shape at worse than median seeing conditions, $M(r,\theta)$ can be written the same way with a different spread, $\sigma_2$. We know that, convolution of two Gaussians are also a Gaussian and using associative property of convolution, we can conclude that $M(r,\theta)$ can be obtained by convolving $\widetilde{M}(r,\theta)$ with a Gaussian function with a spread $\sigma$ and a simple calculation shows that, $\sigma=\sqrt{\sigma_2^2-\sigma_1^2}$.

In real data, the Gaussian function that is needed for convolution to obtain broadened PSF shapes due to seeing is not known a priori. We need to obtain a matching Gaussian kernel using the two star images at different seeing whose inherent PSF shape is the same. To do this we make use of the fact that, convolution operation in real space becomes simple multiplication operation in Fourier space. Consider two background subtracted star images $\tilde{D_1}(r,\theta)$ at median seeing and $\tilde{D_2}(r,\theta)$ with worse seeing condition. We assume that both stars are observed at the same CCD locations. Hence, their inherent PSF shape is same and we can write $\tilde{D_2}(r,\theta)$ as a convolved image of $\tilde{D_1}(r,\theta)$ and another Gaussian function which can be expressed in Fourier space as,
\begin{equation}
    \mathcal{F}[\tilde{D_2}(r,\theta)] = \mathcal{F}[\tilde{D_1}(r,\theta)]\,\times \mathcal{F}[G(r,\theta;\sigma)],
\end{equation}
where, $\mathcal{F}[*]$ denotes the Fourier transform operation. Rearranging the terms and taking inverse Fourier transform we obtain the Gaussian function as \citep{Gordon_2008, Aniano_2011},
\begin{equation}
    G(r,\theta;\sigma) = \mathcal{F}^{-1}\left[ \frac{\mathcal{F}[\tilde{D_1}(r,\theta)]}{W_h(\omega) \,\mathcal{F}[ \tilde{D_2}(r,\theta)]} \right]   \label{eq:mod_inverse_for}
\end{equation}
where, $W_h(\omega)$ is generally a window function in Fourier space ($\omega$) of form,
\begin{equation*}
    W_h(\omega) = \begin{cases} f(\omega,\omega_0) & , \omega \leq \omega_0 \\
                                0   & ,   \omega > \omega_0
                    \end{cases}
 \end{equation*}
 
The Fourier transform of $\tilde D_2(r,\theta)$ in the denominator introduces numeric noise at higher spatial frequencies. Therefore, we avoid high frequencies by using the weight function $W_h$ that acts like a low pass filter and depends on functional form of $f(\omega,\omega_0)$. We used the kernel matching function from the Photutils\footnote{\href{https://photutils.readthedocs.io/en/stable/}{https://photutils.readthedocs.io/en/stable/}} package, that follows Eq. \eqref{eq:mod_inverse_for}. This function takes the PSF shape at median seeing as a reference and, then estimates the associated Gaussian function and its spread for all the broadened PSF shapes at varying seeing conditions. As a low-pass filter in Eq. \eqref{eq:mod_inverse_for}, we used the Tukey window from the Photutils package with the parameter value of 0.5.

Once we obtained the matched Gaussian kernel, we applied Richardson-Lucy de-convolution algorithm on the broadened PSF shape of the stars to obtain the underlying PSF shape expected at median seeing \citep{Richardson_1972, Lucy_1974}. The iterative process of Richardson-Lucy algorithm returns a PSF shape after a fixed number of iterations. This method is highly efficient and quick even for large number of iterations to give desired accuracy in our analysis.

\begin{figure*}[!htbp]
\centering
\includegraphics[clip, trim=1cm .5cm 1.6cm .1cm, width=.9\linewidth]{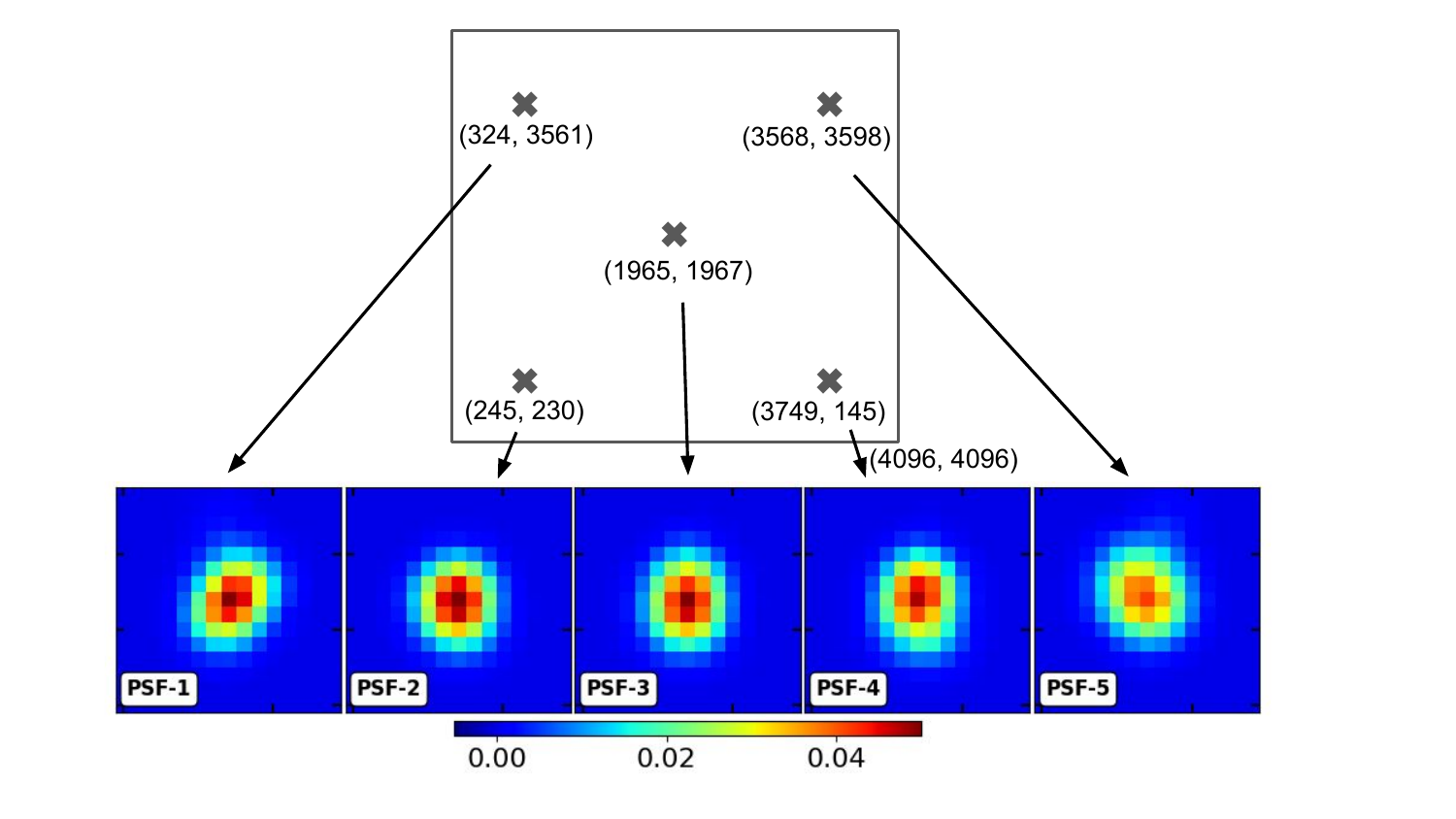}
\caption{Spatial variation of the PSF across the CCD, illustrated at five 
representative locations: the four corners and the centre. PSF-1, PSF-2, 
PSF-3, PSF-4, and PSF-5 correspond to the upper-left, lower-left, central, 
lower-right, and upper-right positions, respectively. Pixel coordinates for each position are provided in parentheses.}
\label{fig:ccd_psf}
\end{figure*}

\section{Simulations} \label{sec:simulations}

We use the Geometric Image Analysis tool in \textsc{Zemax}\footnote{\href{https://www.zemax.com/products/opticstudio}{https://www.zemax.com/products/opticstudio}} to simulate realistic sky observations. These simulations are based on the 1 meter telescope\footnote{\href{https://www.saao.ac.za/astronomers/1-0m/}{https://www.saao.ac.za/astronomers/1-0m/}} specifications at SAAO, the optical model of the WALOP-South instrument, and the local atmospheric seeing conditions \citep{Maharana:2021, Maharana:2024}. The key parameters used in the simulation are listed in \cref{spec_table}. The total number of photons collected at the CCD for a star of given magnitude depends on the spectral transmission profile of the SDSS-$r$ filter, the telescope aperture, and the instrument throughput. To estimate $I_*$ for a star of magnitude $m_*$, we use the relation
\begin{equation}
    I_* = f_{\rm ref}\, A\, w\, t_{\rm exp}\, e\; 
    10^{(m_{\star}/2.512)} \, , \label{eq:4.18}
\end{equation}
where $A$ is the effective collecting area of the telescope, $w$ is the effective filter bandwidth, $t_{\rm exp}$ is the exposure time, $e$ is the throughput of any one of the four instrument channels, and $f_{\rm ref} = 870\,\mathrm{photons\,cm^{-2}\,s^{-1}\,\text{\AA}^{-1}}$ is the reference photon flux for a $0\,\mathrm{mag}$ star. The effective collecting area of the WALOP-South telescope is $7850\,\mathrm{cm}^2$, and the effective bandwidth of the custom SDSS-$r$ filter is $1370\,\text{\AA}$.

\begin{table}
    \centering
    \caption{The key parameters of the WALOP-South optical system. Throughput per channel for telescope plus instrument is the fraction of photons incident on the telescope aperture that are detected by one of the four cameras. }
    \scalebox{0.9}{%
    \large
    \begin{tabular}{c c}
        \hline \hline
         Parameter & Value  \\
         \hline 
         Telescope Aperture & 1~m  \\
         Throughput per Channel & 10 \% \\
         Median Seeing & 1.5\arcs \\
         Filter & SDSS-r \\
         Plate Scale at Detector & 0.51"/pixel \\
         \hline \\
    \end{tabular}
    }
    \label{spec_table}
\end{table}
The mean flux level of the sky background is given by
\begin{equation}
    % I_b=\frac{I_*}{a\,(m_{b}-m_{\star})} \ \label{eq:3.14},
     I_b=\frac{I_*}{a \, 10^{0.4(m_{b}-m_{\star})}} , \label{eq:3.14} 
\end{equation}
where $m_{b}$ is the magnitude of the sky background, $a$ is the pixel area in arcsec$^2$, and $I_b$ is expressed in units of photons\,arcsec$^{-2}$. In all simulated star images, we adopt a sky background level of $m_b = 20\,\mathrm{mag}$.

Using the specifications of the WALOP-South telescope, we generate an input sky image of $4096 \times 4096$ pixels containing multiple stars with Gaussian PSFs of 1.5\arcs\ FWHM at different CCD locations. The simulated star observations are designed to mimic the typical seeing conditions at the WALOP-South observational site. The WALOP-South optical system, in combination with the 
telescope, introduces further distortions to the Gaussian PSF, rendering it extended and irregular. To simulate realistic PSFs, we incorporate the optical, mechanical, and alignment tolerances of the system into the \textsc{Zemax} simulations of WALOP-South, generating PSFs for $10^4$ stars at different CCD locations along with the sky background. We find that the PSF shape varies significantly with CCD location. This variation is illustrated in Fig.~\ref{fig:ccd_psf}, which shows five representative PSFs: four extracted from diagonally opposite positions near the corners of the CCD, and one from near the centre. All subsequent analyses are performed on these five selected PSF shapes.

\begin{figure}[!htbp]
\centering
\includegraphics[width=\columnwidth, keepaspectratio=True]{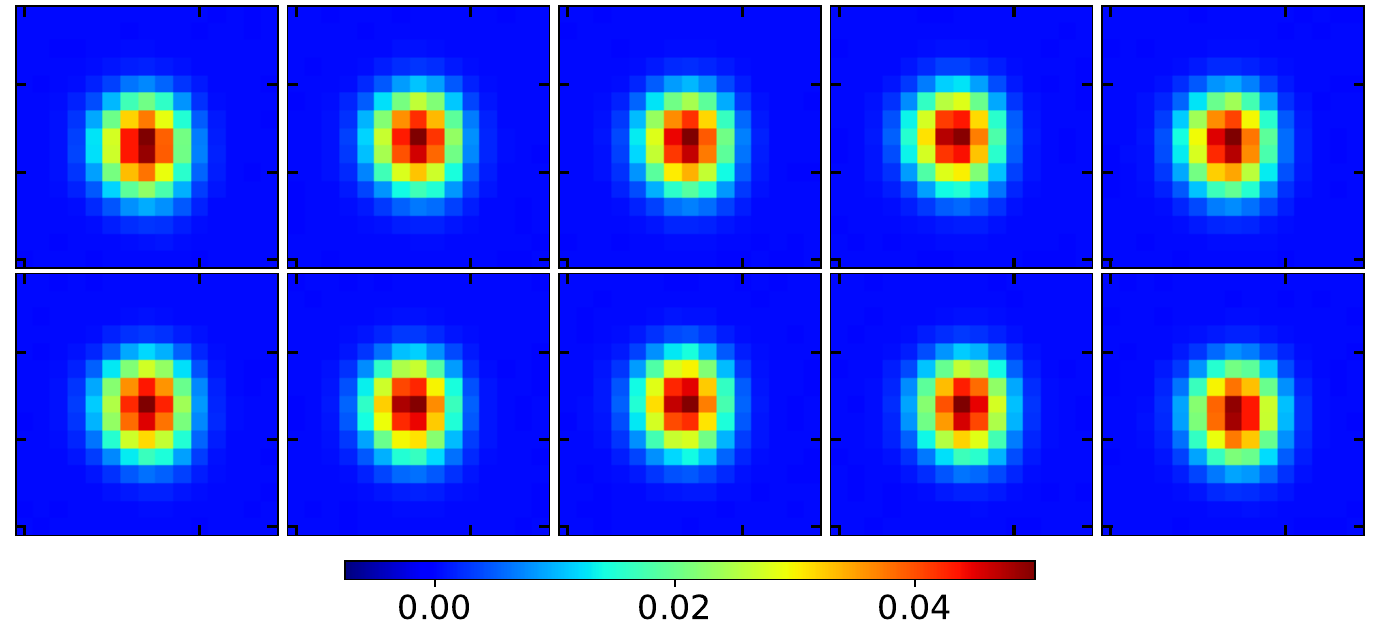}
\caption{Morphological variations across ten realisations of PSF-2, each 
with a random sub-pixel centroid shift within the central pixel. The images 
are displayed as $15 \times 15$ pixel cutouts to highlight the subtle 
structural differences arising from these sub-pixel shifts. }
\label{fig:shifted_psf}
\end{figure}

For each of the five selected PSF shapes, we scale the total photon count to that of a 12\,mag star with a fixed exposure time of 30\,s. We simulate ten independent realisations of each star image by adding Poisson noise to individual pixels. To mimic real observations, we introduce a random sub-pixel shift to the centroid of each realisation using a Fourier-based sub-pixel shifting technique. According to the Fourier shift theorem, a spatial shift in real space corresponds to multiplication by a complex phase factor in Fourier space \citep{Bracewell2000}. By applying the desired sub-pixel shift through phase modulation of the Fourier-transformed image followed by an inverse Fourier transform, one obtains a real-space image accurately shifted at sub-pixel resolution. This technique preserves both the total flux and the PSF shape, while avoiding the resolution loss and interpolation artifacts inherent to real-space shifting methods \citep{Guizar-Sicairos:08}. Figure~\ref{fig:shifted_psf} shows ten independent realisations of PSF-2 with randomly shifted centroids within the central pixel, illustrating the resulting variations in the apparent PSF morphology.

We performed a curve-of-growth analysis on the background-subtracted average star image to determine the square aperture size enclosing 99.9\% of the total photon counts for a $12\,\mathrm{mag}$ star. The curve-of-growth analysis yields a consistent aperture size across different realisations of the star. To maintain consistency with PSFs across the CCD, we adopt a fixed image size of $25\times 25$~pixels for all cases and use this size for subsequent analysis. We normalised all the realisations of the bright star PSF within the square aperture. We then performed PSF modelling and the reconstructed model PSF is subsequently employed to perform PSF photometry on simulated faint $16\,\mathrm{mag}$ target stars, to which Poisson noise has been added to replicate WALOP-South Instrument.

\begin{figure} [!htbp]
\centering
\includegraphics[width=\hsize]{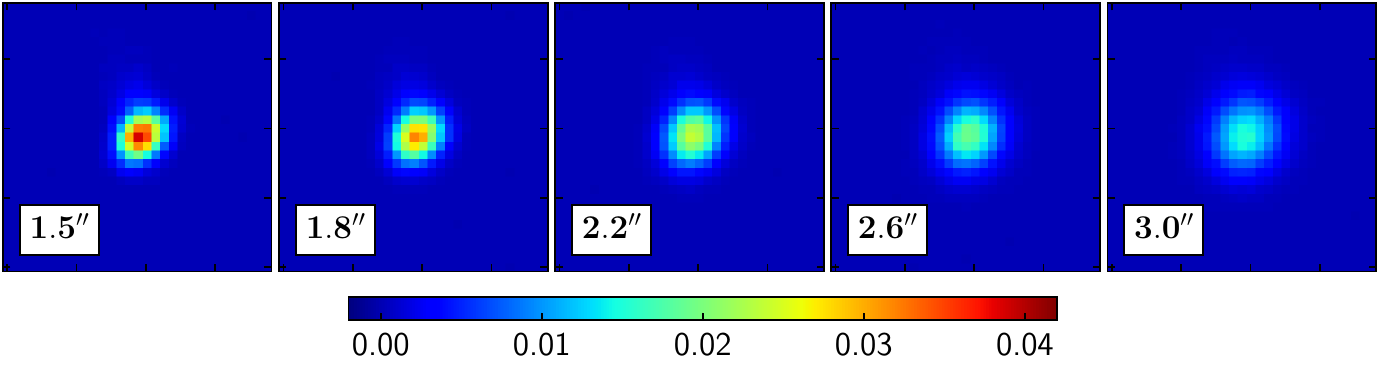}
\caption{One realisation of a $12\,\mathrm{mag}$ bright star for five different 
seeing conditions, ranging from 1.5\arcs\ to 3\arcs\ FWHM.}
\label{fig:seeing_sims}
\end{figure}

The \textsc{Zemax} simulations were produced considering median seeing of 1.5\arcs at the WALOP-South observation site. To perform the analysis under varying seeing conditions, we select PSF-1 as a reference PSF, assuming its centroid fixed at the origin $(x^c, y^c) = (0, 0)$. To simulate different seeing conditions worse than the median seeing of 1.5\arcs\ FWHM, we convolved the reference PSF with Gaussian kernels of appropriate standard deviation, generating ten PSF shapes corresponding to seeing conditions ranging from 1.8\arcs\ to 3\arcs\ FWHM in steps of 0.2\arcs. Each of these ten PSF shapes is scaled separately to simulate $12\,\mathrm{mag}$ bright stars with an exposure time of $30\,\mathrm{s}$. A sky background of $20\,\mathrm{mag}$ and Poisson noise are added to all simulations. A single realisation of a bright star for five different seeing conditions are shown in Fig.~\ref{fig:seeing_sims}. At each seeing condition, the ten realisations of background-subtracted bright stars are used to characterise the seeing-induced PSF broadening and to reconstruct the model PSF. PSF photometry is then applied to $10^4$ realisations of the faint stars.

We define three criteria to validate the PSF modelling and photometry 
pipeline:
\begin{enumerate}
    \item The total photon counts of the reconstructed model PSF is close to unity.
    \item The mean photometric result from $10^4$ faint stars realisations agrees with 
    the input total photon count to within $1\sigma$.
    \item The correlation between the input star image and the residual image 
    after PSF photometry is negligible.
\end{enumerate}
All three criteria must be simultaneously satisfied for the PSF modelling and PSF photometry pipeline to be considered statistically unbiased and suitable for subsequent scientific analysis.

\section{Results}\label{sec:results}

In this section, we determine the scale parameter $\beta$, the number of shapelet coefficients $n_{\rm max}$, and the exposure time $\texp$ required to achieve the target photometric sensitivity of $0.15\%$ for realistic sky observations from the WALOP-South telescope. The scale parameter $\beta$ characterises the mean PSF broadening across different CCD locations. We first fit for $\beta$ using $10^4$ stars at different CCD locations, and adopt the best-fit value for all subsequent analyses. We then determine the minimum number of shapelet coefficients $n_{\rm max}$ required to accurately model the five selected PSFs, subject to the constraint that each model PSF is normalised to unity. With $n_{\rm max}$ fixed, we determine the exposure time $\texp$ needed to meet the photometric sensitivity requirement of the \pasiphae survey. Each of these steps is discussed in detail in the following subsections.

\begin{figure}
\centering
\includegraphics[width=\hsize]{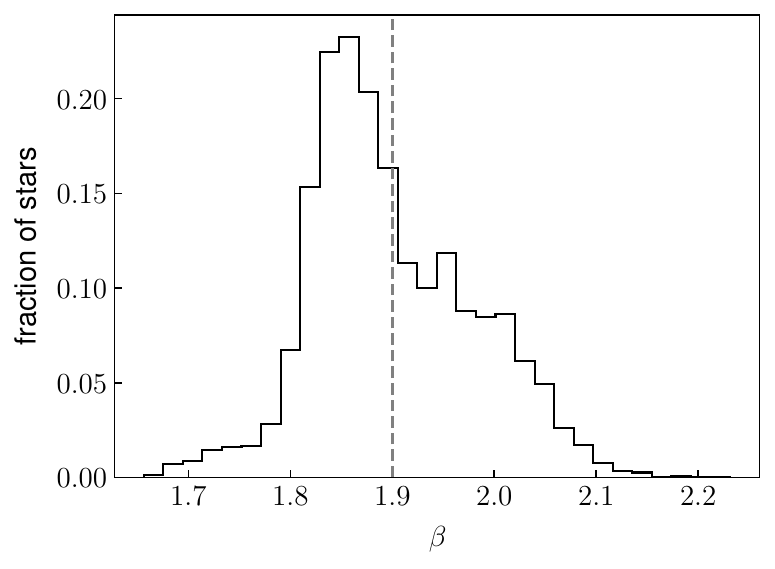}
\caption{The histogram of best-fit $\beta$ values for $10^4$ stars distributed over the whole CCD. The grey vertical line represents the median of the distribution corresponding to $1.9$ pixels.} 
\label{fig:fix_beta} 
\end{figure}

\begin{figure*}
\centering
\includegraphics[width=0.9\hsize]{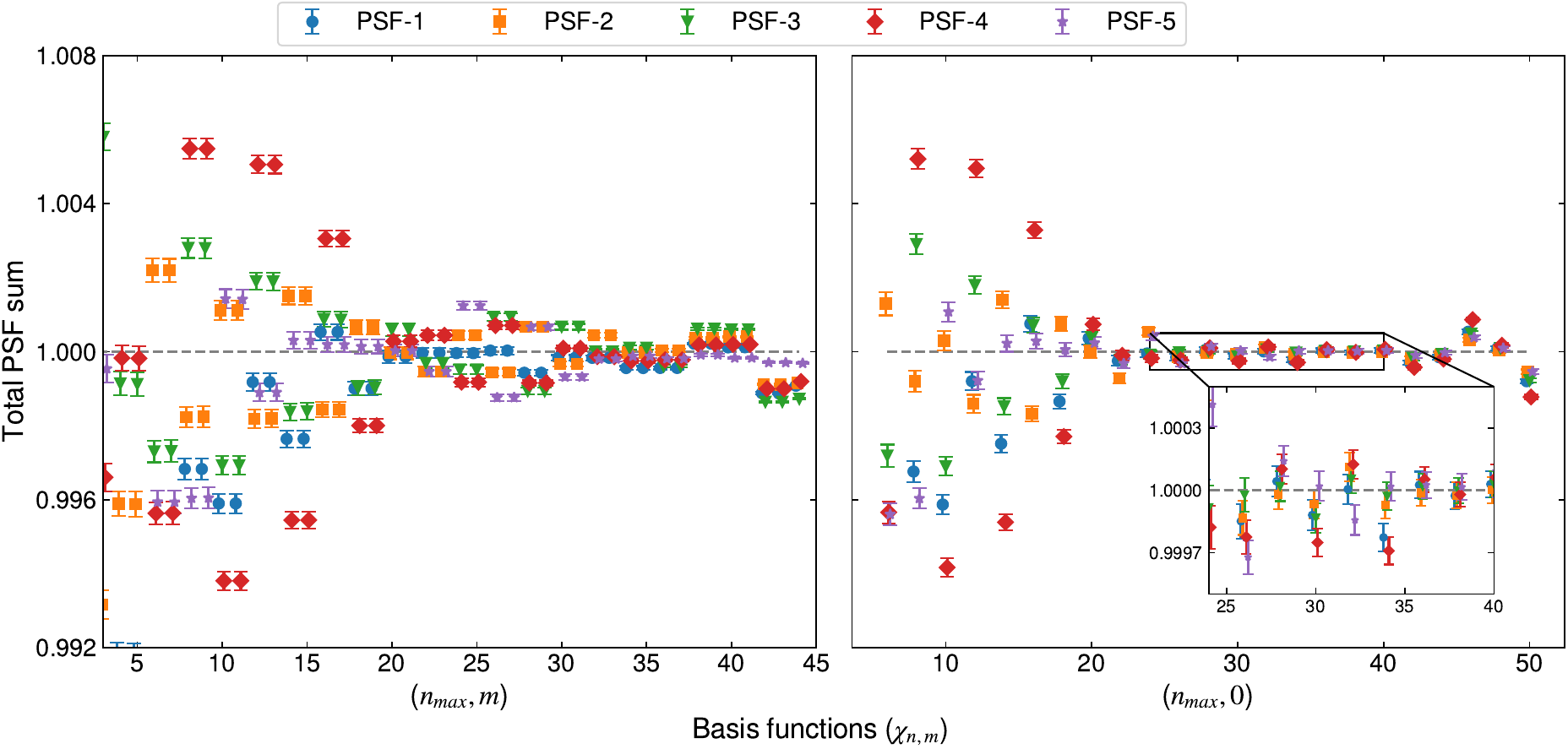}
\caption{Determination of the required number of shapelet coefficients to model the five PSFs. \textit{Left panel}: Normalised PSF sum as a function of $n_{\rm max}$ for PSF modelling using ten realisations of a background-subtracted $12\,\mathrm{mag}$  star image. \textit{Right panel}: Normalised PSF sum as a function of the number of central ($m = 0$) shapelet terms for PSF modelling using the same ten realisations. The dashed grey line indicates unity.} 
\label{fig:ps_cons}
\end{figure*}

\subsection{Fixing scale parameter $\beta$}

We first constrain the scale parameter $\beta$ defined in Eq.~\eqref{eq:3.1}. The zeroth-order shapelet basis function 
($n = 0$, $m = 0$) is proportional to a 2D symmetric Gaussian 
\citep{Massey:2005},
\begin{equation}
    \chi_{00}(r,\theta;\beta) \propto \exp\left(-\frac{r^2}{2\beta^2}\right) 
    \, , \label{eq:6.1.1}
\end{equation}
where $\beta$ is equivalent to the standard deviation $\sigma$ of the Gaussian. We fit this zeroth-order basis function to each bright star cutout at different CCD locations to obtain the best-fit value of $\beta$. This is equivalent to first approximating the PSF as a 2D symmetric Gaussian, with higher-order shapelet coefficients subsequently used to model the residual features of the PSF. Fig.~\ref{fig:fix_beta} shows the distribution of best-fit $\beta$ values for $10^4$ stars across the CCD. The distribution is skewed, reflecting the asymmetric spatial distribution of stars across the CCD plane. The median value is $\beta = 1.9$ pixels, with a standard deviation of $0.1$ pixels, corresponding to 2.28\arcs\ FWHM on the sky. The additional PSF broadening is introduced by the instrument optics. We therefore adopt $\beta = 1.9$ pixels as a fixed parameter for all subsequent photometric analyses.

\subsection{Fixing the value of $n_{\rm max}$}{\label{sec:5.2}}

Following the first criterion (see Sect.~\ref{sec:simulations}), the total photon counts of the reconstructed model PSF using the polar shapelets formalism must approach unity. To model the PSF, we use the background-subtracted, normalised image of a bright star and progressively include shapelet coefficients up to order $n_{\rm max}$, starting from $n_{\rm max} = 1$. With a small number of shapelet coefficients, the model fails to fully capture the PSF morphology, resulting in a significant bias in the total sum of the model PSF, as expected. Fig.~\ref{fig:ps_cons} shows that the total sum of the PSF approaches unity as $n_{\rm max}$ increases. Including a sufficient number of shapelet coefficients, the total sum converges to unity with a deviation of less than $0.1\%$, satisfying the first criterion. However, increasing $n_{\rm max}$ beyond this value causes the sum to deviate from unity once more, indicative of PSF overfitting.  The associated uncertainties shown in Fig. \ref{fig:ps_cons} are estimated from $10^4$ realisations of the simulated source image.

When including all shapelet coefficients up to a given $n_{\rm max}$, we observe a pairing of consecutive values in the left panel of Fig.~\ref{fig:ps_cons}, indicating that adding successive pairs of coefficients yields diminishing improvement in the total photon count convergence. This behaviour arises because, at every alternate $n_{\rm max}$, a circularly symmetric term is added to the basis set that contributes meaningfully to the model PSF, while the intermediate non-symmetric term provides negligible improvement. Motivated by this observation, we perform a complementary simulation in which, after including all terms up to $n_{\rm max} = 5$, only the central (circularly symmetric) shapelet coefficients are included for higher orders. We find that retaining only the central shapelets at each order is as effective as including the full set of coefficients. Convergence is first achieved at the central term $\chi_{10,0}$, with stable convergence attained around $\chi_{20,0}$. The right panel of Fig.~\ref{fig:ps_cons} shows the convergence of the normalised PSF integrated count as a function of the number of central terms included. For all subsequent analyses, we adopt a basis comprising all shapelet coefficients up to $n_{\rm max} = 5$, supplemented by central terms up to $\chi_{28,0}$, yielding a total of 33 shapelet coefficients. 
This compact representation accurately captures the PSF morphology while significantly reducing the computational cost of the algorithm. PSF modelling with sub-pixel shift typically requires few seconds of CPU time on a single core.

\begin{figure*}
\centering
\includegraphics[width=0.85\hsize]{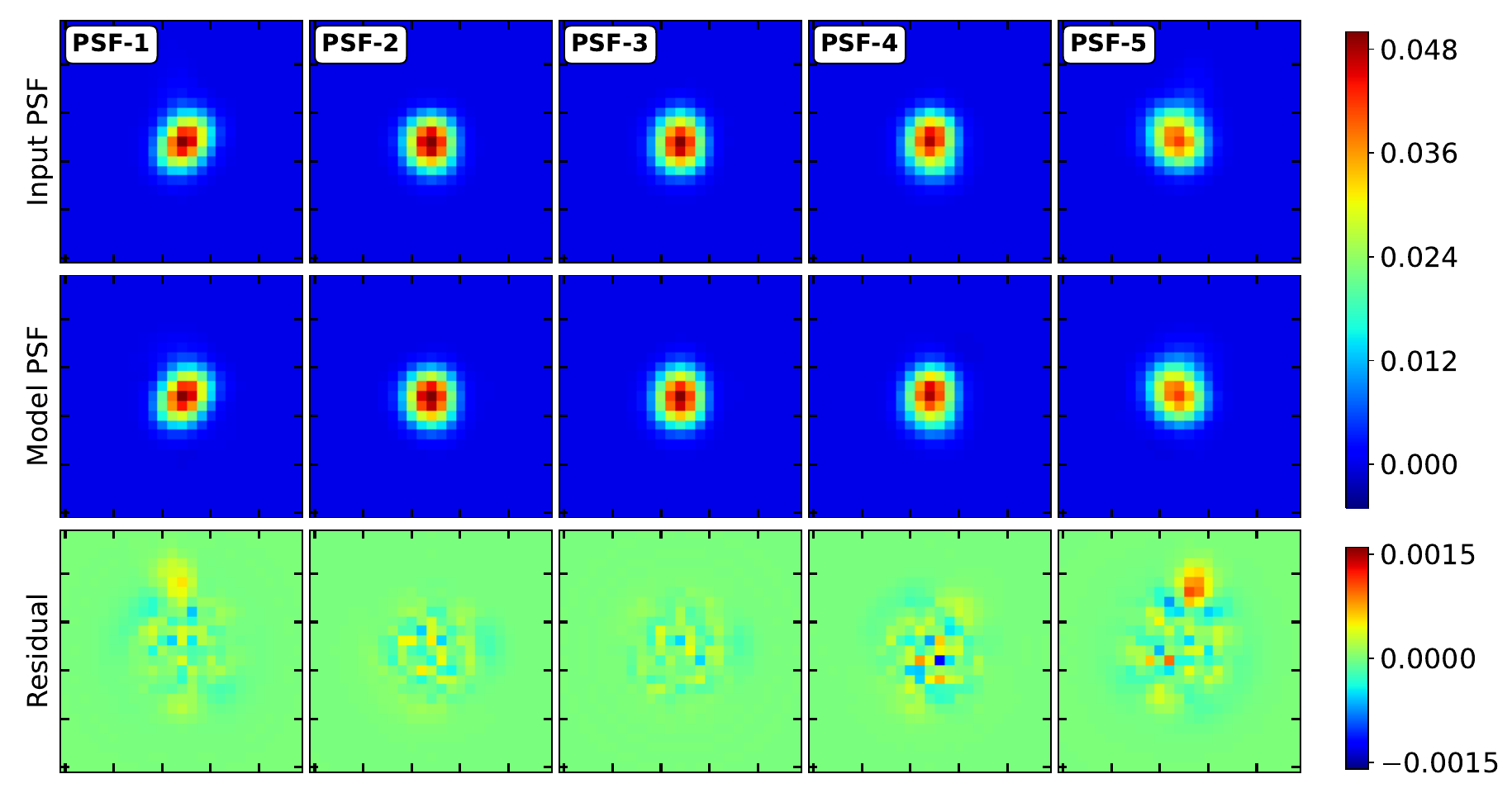}
\caption{PSF model for stars at five CCD locations. \textit{Top panel}: normalised 
mean of ten realisations of a background-subtracted $12\,\mathrm{mag}$  star image, 
representing the observed PSF at each CCD location. \textit{Middle panel}: 
reconstructed model PSF using 33 shapelet coefficients. \textit{Bottom 
panel}: residual map between the observed PSF and reconstructed model PSF. } 
\label{fig:psf_model}%
\end{figure*}

The top two panels of Fig.~\ref{fig:psf_model} show the normalised, background-subtracted PSF of a $12\,\mathrm{mag}$ bright star alongside the corresponding model PSF for five selected CCD locations. Although the recovered PSFs are not perfectly centred owing to the sub-pixel shifts introduced across realisations, these centroid offsets do not affect the photometric precision, since the value of $I_\star$ and the centroid coordinates $(x^c, y^c)$ are treated as independent free parameters in the minimisation of Eq.~(\ref{eq:3.10}). The specific sub-pixel shifts adopted for ten realisations are listed in Table~\ref{tab:sp_shift} for PSF-2. The bottom panel of Fig.~\ref{fig:psf_model} shows the residual map after correcting for the centroid offsets. The residuals are randomly distributed around zero, with peak fluctuations below $6\%$ of the maximum pixel intensity of the normalised PSF. This level of convergence confirms that the polar shapelets formalism accurately reconstructs complex PSF morphologies across the full CCD field-of-view.

\begin{figure}
\centering
\includegraphics[width=0.6\hsize]{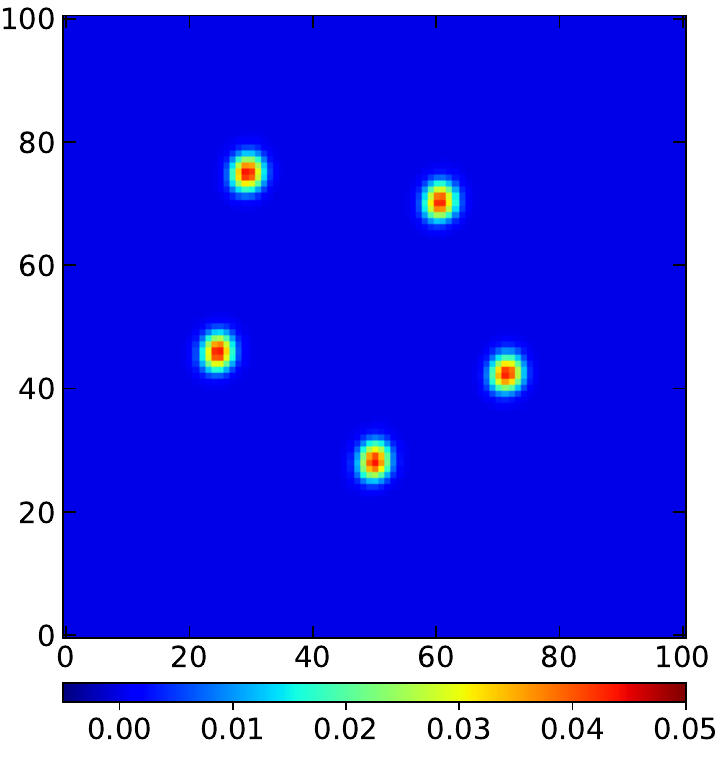}
\caption{\textsc{Zemax}-generated output field containing five stars randomly positioned across the field. Each star is normalized and no background is included.}
\label{fig:zemax_field}
\end{figure}

Instead of manually introducing sub-pixel shifts to the star images, we also test our algorithm directly on a simulated \textsc{Zemax} output containing multiple stars. We use a $101 \times 101$ pixel \textsc{Zemax}-generated image with five stars randomly distributed across the field, with no sky background added. The simulated field for five stars using \textsc{Zemax} software is shown in Fig.~\ref{fig:zemax_field}.

\begin{figure*}
\centering
\includegraphics[width=0.85\hsize]{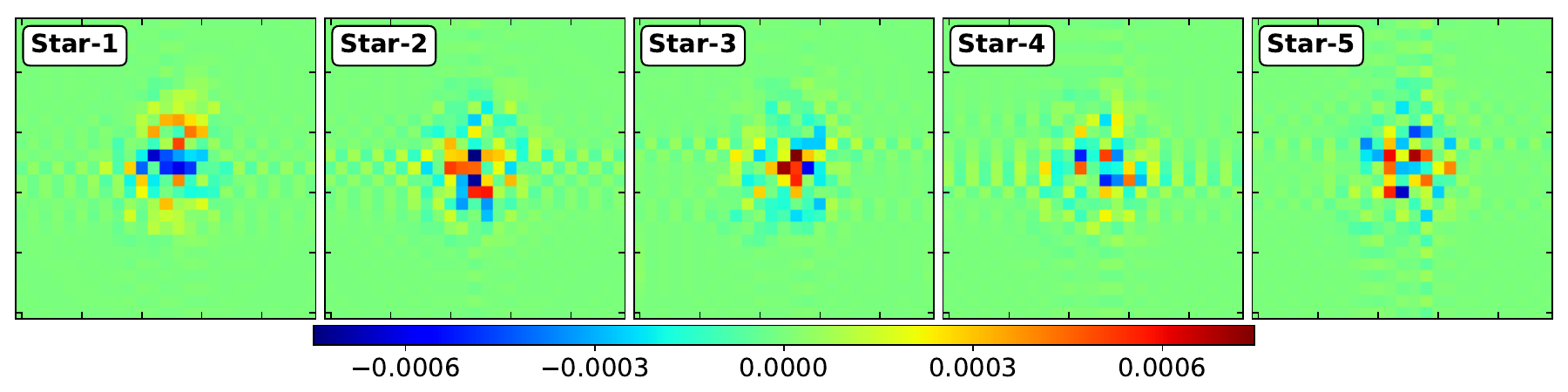}
\caption{Residual maps obtained after subtracting the model PSF from the five input star PSFs, showing no significant systematic correlations.}
\label{fig:zemax_residue}
\end{figure*}

Star positions are identified using the DAOStarFinder algorithm, provided by the Photutils package. For each detected star, a $25 \times 25$ pixel cutout centred on the star position is extracted and passed as input to the PSF modelling pipeline, using 33 shapelet coefficients. The total integrated photon count of the reconstructed model PSF converges to unity, satisfying the first normalisation criterion (see Sect.~\ref{sec:simulations}). The residual maps between the input star images and the model PSF, presented in Fig.~\ref{fig:zemax_residue}, show no discernible systematic patterns or correlated structures, confirming that the PSF model successfully captures the intrinsic morphological variations across the field and satisfying the third modelling criterion.

\begin{table}
    \centering
    \caption{Recovery of input centroid shifts ($x^c, y^c$) for PSF-2, derived from ten realisations of a $12\,\mathrm{mag}$ star image. Columns 2 and 3 list the input random shifts introduced to the centroid positions. Columns 4 and 5 give the recovered shifts obtained from the PSF model, with the differences shown in columns 6 and 7. The consistent residuals across all realisations yield mean differences of $(0.036,\, 0.073)$, corresponding to the intrinsic relative shift between the constructed model PSF and the initial \textsc{Zemax} output PSF. All shifts are quoted in pixel units.}    
    \scalebox{0.8}{%
    \large
    \begin{tabular}{ccccccc}
        \hline \hline &  & \\[-0.2cm]
          Realisation  &   \multicolumn{2}{c}{input} & \multicolumn{2}{c}{recovered} & \multicolumn{2}{c}{difference} \\
          \cline{2-7}\\[-0.1cm]
            &   $x^c$ & $y^c$ &  $x^c$ & $y^c$ &  $\Delta x^c$ & $\Delta y^c$ \\
         \hline &  & \\[-0.2cm]
         1 & -0.051 & -0.379 &  -0.015 & -0.305 & 0.036 & 0.074\\
         2 & 0.039  & -0.052 & 0.073 & 0.020 & 0.034 & 0.072\\
         3 & -0.064 & -0.136 & -0.032 & -0.055 & 0.031 & 0.081\\
         4 & -0.236 & 0.095 & -0.200 & 0.167 & 0.036 & 0.071\\
         5 & -0.160 & -0.187 & -0.118 & -0.112 & 0.042 & 0.075\\
         6 & 0.097  & 0.023 & 0.131 & 0.094 & 0.034 & 0.071 \\
         7 & -0.292 & 0.011 & -0.254 & 0.090 & 0.038 & 0.079\\
         8 & -0.252 & 0.228 & -0.214 & 0.298 & 0.039 & 0.070 \\
         9 & 0.283  & 0.004 & 0.316 & 0.067 & 0.032 & 0.071\\
         10 & 0.277 & -0.336 & 0.312 & -0.267 & 0.036 & 0.070\\
         \hline
    \end{tabular}
    }
    \label{tab:sp_shift}
\end{table}

\begin{figure}
\centering
\includegraphics[width=0.95\hsize]{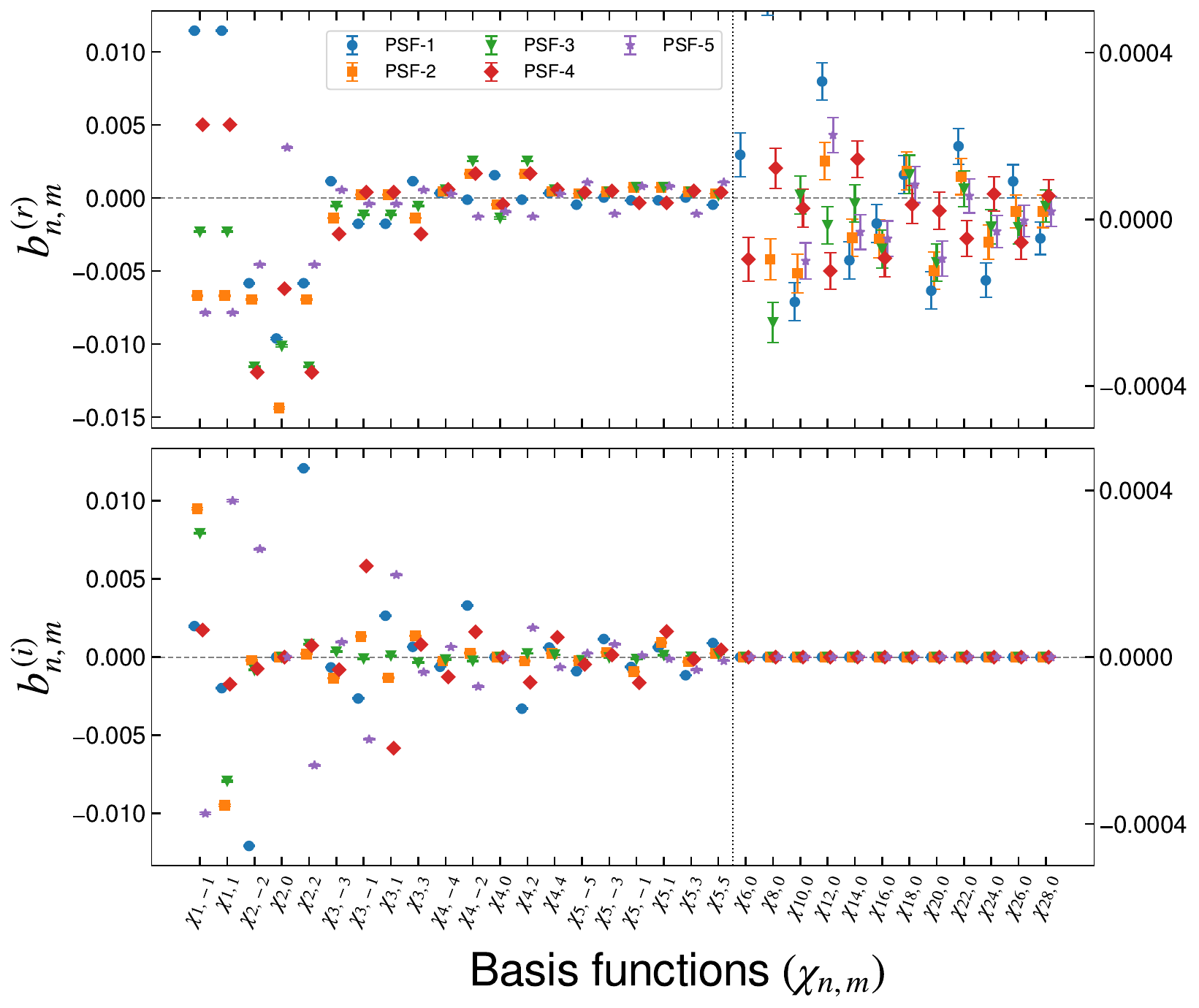}
\caption{Real (\textit{top panel}) and imaginary (\textit{bottom panel}) parts of the shapelet coefficients $b_{n,m}$, with associated error bars derived from $10^4$ independent realisations. All polar shapelet terms up to $n_{\rm max} = 5$ 
(excluding $\chi_{0,0}$) are shown, followed by the higher-order central terms up to $\chi_{28,0}$. The dashed horizontal line denotes the zero baseline. The coefficients for $n_{\rm max} > 5$ are highlighted with a zoomed-in view.}
\label{fig:bnm_err}
\end{figure}

\subsection{Uncertainties in coefficients}

We estimate the $1\sigma$ uncertainties in the shapelet coefficients by computing the standard deviation across $10^4$ independent realisations of the PSF modelling. Figure~\ref{fig:bnm_err} presents the measured coefficients alongside their $1\sigma$ uncertainties, allowing us to assess the statistical significance of higher-order shapelet coefficients in the reconstructed model PSF.

All shapelet basis coefficients exhibit uncertainties significantly smaller than their corresponding mean values, indicating that these coefficients are robustly constrained by the data. The real components, $b_{n,m}^{(r)}$, shown in the upper panel of Fig.~\ref{fig:bnm_err}, exhibit a clear convergence toward zero around the $\chi_{24,0}$ mode, beyond which the coefficients enter a noise-dominated regime. The comparatively larger coefficient amplitudes at low $n_{\rm max}$ confirm that the lowest-order basis functions capture the dominant structural features of the PSF, while higher-order basis functions primarily encode small-scale noise and residual modelling imperfections.

\subsection{Fixing $\texp$ for target stars}

\begin{figure}
\centering
\includegraphics[width=\hsize]{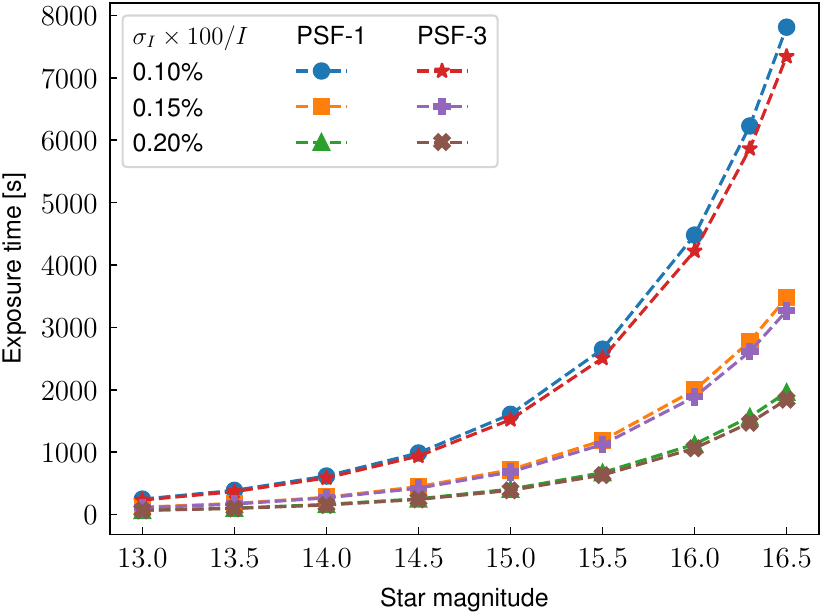}
\caption{Average exposure time required to achieve a fixed accuracy in photometry for two different locations on CCD and different star magnitudes.}
\label{fig:exp_time_mag}
\end{figure}

The photometric accuracy achieved by PSF photometry depends on both parameters: $m_*$ and $\texp$. Achieving a fixed level of accuracy requires longer exposure times for faint target stars, reflecting the exponential decrease in photon counts with increasing magnitude. We apply the PSF model reconstructed from $12\,\mathrm{mag}$ star observations to perform PSF photometry on multiple realisations of fainter target stars over a range of exposure times, and identify the minimum exposure time required to reach a photometric accuracy of $0.15\%$. Fig.~\ref{fig:exp_time_mag} shows the exposure time required for stars of various magnitudes to achieve different levels of photometric accuracy, for two representative PSFs selected from Fig.~\ref{fig:ccd_psf}. For the target star magnitude of $m_\star = 16$, we adopt an exposure time of $2010\,\mathrm{s}$ ($\approx 32.5\,\mathrm{min}$) to achieve the required photometric accuracy of $0.15\%$.

\subsection{PSF photometry on target stars}      

To validate the reconstructed model PSF, we conduct Monte Carlo simulations at five CCD locations. We generate $10^4$ realisations of a $16\,\mathrm{mag}$ star embedded in a sky background of $20\,\mathrm{mag}$ with an exposure time of $2010\,\mathrm{s}$, drawing photon counts from a Poisson distribution to replicate realistic detector noise. The analysis is performed at the same five CCD positions at which the PSF variations were characterised (see Fig.~\ref{fig:ccd_psf}). To isolate the photometric precision from potential centroid alignment errors, the model PSF centroids are registered to the simulated star coordinates prior to the extraction of total photon count. This ensures that the resulting photometric accuracy reflects the fidelity of the model PSF reconstruction rather than sub-pixel centroid offsets.

We estimate $I_*$ and sub-pixel centroid position by minimising the loss function $S$ defined in Eq.~(\ref{eq:3.10}). We employ model-weighted least squares, in which the variance $\sigma^2(r,\theta)$ is set proportional to the model-predicted intensity. While uniform weighting is adopted during the initial centroiding step for numerical stability, model weighting is the statistically optimal approach for flux estimation in the Poisson-limited regime \citep{Mighell:2005}, as it priorities high-flux pixels while suppressing background fluctuations \citep{Stetson:1987}.

\begin{figure}
\centering
\includegraphics[width=\hsize]{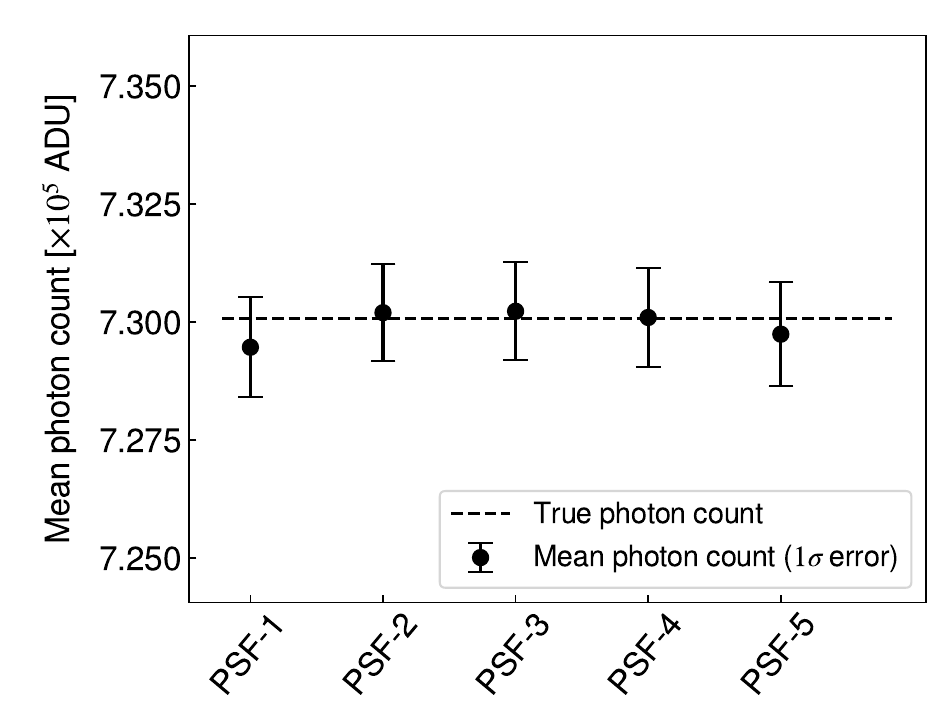}
\caption{Distribution of the mean recovered photon count (in units of $10^5$ ADU) from PSF photometry 
for five PSF shapes, derived from $10^4$ realisations of a $16\,\mathrm{mag}$ 
star with an exposure time of $2010\,\mathrm{s}$.}
\label{fig:psf_phot}
\end{figure}

The resulting distributions of the recovered photon counts are compared against the true input values in Fig.~\ref{fig:psf_phot}. The value of $I_*$ is successfully recovered at all five CCD locations with a photometric accuracy of $0.15\%$, confirming that the adopted parameter values satisfy the criteria  outlined in Sect.~\ref{sec:simulations}.

\subsection{PSF photometry versus aperture photometry}

It is well established that PSF photometry outperforms aperture photometry, particularly in optical systems characterised by significant spatial PSF distortions \citep{Stetson:1987}. To quantify this advantage for the WALOP-South instrument, we perform a comparative analysis between the two approaches. A model PSF is reconstructed from multiple realisations of a $12\,\mathrm{mag}$ star using the polar shapelets formalism, and subsequently used to perform PSF photometry on a single realisation of a $16\,\mathrm{mag}$ faint target star embedded in a sky background of $20\,\mathrm{mag}$. Aperture photometry is performed on the same realisation, with the optimal aperture radius determined via a curve-of-growth analysis enclosing $99.9\%$ of the total flux.

The results of this comparison are summarised in \cref{tab:psf_vs_aper}. PSF photometry consistently yields flux estimates closer to the true photon count and achieves lower photometric uncertainty than aperture photometry across all five PSF cases. The improvement is most pronounced for PSF-1 and PSF-5, which exhibit the greatest spatial distortions, as expected from the sensitivity of PSF photometry over aperture photometry.

Given the highly distorted PSF shapes characteristic of the WALOP-South optical system, our results strongly favor the PSF photometry approach. This method delivers a stable photometric accuracy of approximately $0.15\%$, largely independent of PSF shape, making it the preferred choice for precision photometry for the WALOP-South telescope. A performance comparison of PSF photometry versus aperture photometry for stellar observations from the RoboPol survey \citep{Ramaprakash:2019} is presented in Appendix~\ref{appendix:B}.

\begin{table}
    \centering
    \caption{Comparison of PSF photometry and aperture photometry results for a single 
realisation of a $16\,\mathrm{mag}$  faint star. The true photon count is 
$730{,}068\,\mathrm{ADU}$. }    
    \scalebox{0.9}{%
    \large
    \begin{tabular}{ccc}
        \hline \hline &  & \\[-0.2cm]
            &   PSF photometry counts & Aperture photometry counts \\
            &   (in ADU)  &  (in ADU)  \\
         \hline &  & \\[-0.2cm]
         PSF-1 & 729457 $\pm$ 1071 & 728702 $\pm$ 1476 \\
         PSF-2 & 730213 $\pm$ 1027 & 730617 $\pm$ 1199 \\
         PSF-3 & 730233 $\pm$ 1035 & 730212 $\pm$ 1199 \\
         PSF-4 & 730111 $\pm$ 1041 & 730461 $\pm$ 1199 \\
         PSF-5 & 729732 $\pm$ 1098 & 730461 $\pm$ 1678 \\
         \hline
    \end{tabular}
    }
    \label{tab:psf_vs_aper}
\end{table}

\subsection{PSF photometry in varying seeing conditions}

The output spread values $\sigma_{\rm out}$ derived from the PSF-matching algorithm are compared with the true input values $\sigma_{\rm in}$ used to simulate varying seeing conditions, with agreement found to within $0.05\%$. The recovered standard deviations of the matching Gaussian kernel are highly accurate; however, their accuracy depends on the window function used to suppress high-frequency noise in the algorithm. The cut-off frequency is chosen such that the recovered values agree with the true values to within $0.1\%$.

\begin{figure}
\centering
\includegraphics[width=\hsize]{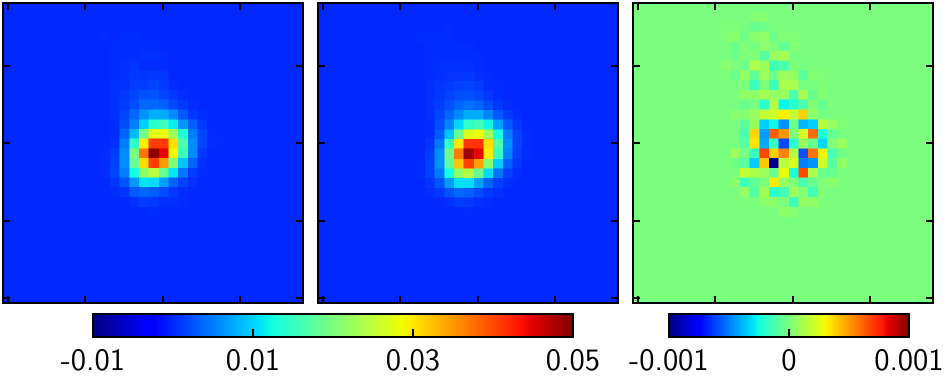}
\caption{(\textit{Left panel}) Input PSF at median seeing. (\textit{Middle panel}) Mean deconvolved PSF under degraded seeing conditions. (\textit{Right panel}) Residuals between the two PSFs.}
\label{fig:decon_psf}
\end{figure}

\begin{figure}
\centering
\includegraphics[width=0.8\hsize]{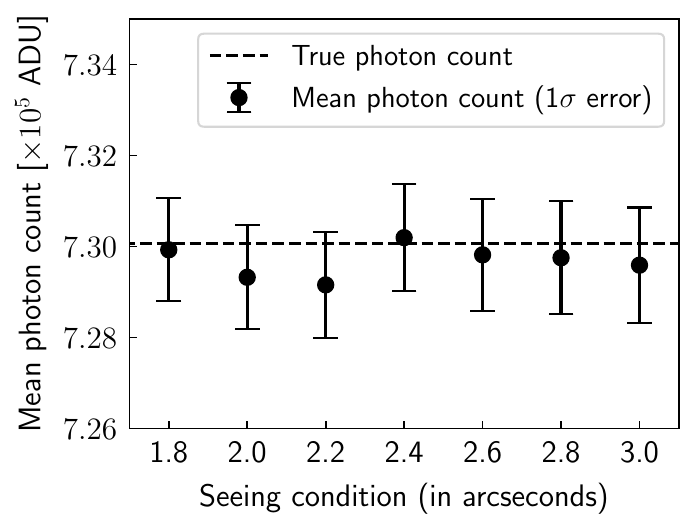}
\caption{Mean flux recovered via PSF photometry from $10^4$ independent realisations of faint target stars under varying seeing conditions.}
\label{fig:phot_seeing}
\end{figure}

Using the derived Gaussian matching kernels, all simulated star images under varying seeing conditions are deconvolved. The mean of the deconvolved PSF shapes is computed and found to be consistent with the reference PSF at 1.5\arcs\ FWHM seeing, as expected from the simulation setup. However, owing to numerical noise accumulated across the processing steps, low-level residuals are present between the two PSFs, as shown in Fig.~\ref{fig:decon_psf}. The reference PSF recovered by deconvolution is subsequently modelled using shapelet coefficients. For each seeing condition, the reconstructed model PSF is convolved with a Gaussian of the corresponding derived standard deviation $\sigma_{\rm out}$, yielding an effective PSF model tailored to that condition. PSF photometry is then performed on the respective simulated star images using these seeing-dependent PSF models. For the photometric analysis, all simulated star images are scaled to $16\,\mathrm{mag}$ with a sky background of $20\,\mathrm{mag}$ and an exposure time of $2010\,\mathrm{s}$, with Poisson noise included. The PSF photometry results for $10^4$ faint stars under each of the seven seeing conditions are shown in Fig.~\ref{fig:phot_seeing}. The input flux is recovered to within one standard deviation in all cases.

\section{Conclusions}\label{sec:conclusion}

In this paper, PSF photometry is implemented on simulated images from the WALOP-South instrument. It is demonstrated that polar shapelets provide a natural basis for PSF modelling and photometry in wide-field surveys such as the \pasiphae\ survey. The PSF modelling and photometry analysis are carried 
out at five different CCD positions, some of which exhibit highly distorted PSF profiles.

Two practical challenges are addressed: sub-pixel shifts in the centroids of reference stars, and variations in seeing conditions. In both cases, the targeted photometric accuracy of $0.15\%$ is achieved using only 33 shapelet coefficients, demonstrating the computational efficiency of the 
method. The analysis shows that retaining only the central terms of higher-order shapelets basis ($n_{\rm max} > 5$) is sufficient to reach the required accuracy. Pre-computing and storing the principal polar shapelets as images at the required resolution further improves efficiency when modelling multiple PSFs. If the PSF shape varies smoothly across the detector, coefficient maps can be constructed for all principal polar shapelets used in the modelling. Such continuous maps would enable photometry to be performed at arbitrary detector positions.

The PSF photometry method is compared with the widely used aperture photometry approach. PSF photometry yields consistent photometric accuracy across all five selected PSF shapes without bias in the mean photon count, whereas the performance of aperture photometry is found to be strongly dependent on the PSF shapes. The analysis further provides constraints on $\SNR$ and $\texp$ for observations of both bright and faint stars required to achieve the target photometric accuracy. However, these predictions are sensitive to observing conditions, including atmospheric broadening of the PSF and the pixel response function of the CCD. In actual observations, the same analysis will be applied to determine optimal parameters such as the scale parameter $\beta$, which characterises the spatial extent of the PSF, and the number of shapelet coefficients required for modelling.

\begin{acknowledgements}

The \pasiphae program is supported by grants from the European Research Council (ERC) under grant agreements No 771282 and No 772253; by the National Science Foundation (NSF) award AST-2109127; by the National Research Foundation of South Africa under the National Equipment Programme; by the Stavros Niarchos Foundation under grant \pasiphae; and by the Infosys Foundation. S.K,  J.A.K., and N.M. were funded by the European Union ERC-2022-STG - BOOTES - 101076343. D.B. acknowledges support from the European Research Council (ERC)  under the Horizon ERC Grants 2021 program under grant agreement No. 101040021. Views and opinions expressed are however those of the author(s) only and do not necessarily reflect those of the European Union or the European Research Council Executive Agency. Neither the European Union nor the granting authority can be held responsible for them. This research made use of Photutils, an Astropy package for detection and photometry of astronomical sources \citep{larry_bradley_2020_4044744}. Some part of image analysis was made easy by Sci Kit image python module \citep{van2014scikit}. 

\end{acknowledgements}

\bibliographystyle{aa}
\bibliography{bibliography}

@misc{Tassis:2018,
      title={PASIPHAE: A high-Galactic-latitude, high-accuracy optopolarimetric survey}, 
      author={Konstantinos Tassis and Anamparambu N. Ramaprakash and Anthony C. S. Readhead and Stephen B. Potter and Ingunn K. Wehus and Georgia V. Panopoulou and Dmitry Blinov and Hans Kristian Eriksen and Brandon Hensley and Ata Karakci and John A. Kypriotakis and Siddharth Maharana and Evangelia Ntormousi and Vasiliki Pavlidou and Timothy J. Pearson and Raphael Skalidis},
      year={2018},
      eprint={1810.05652},
      archivePrefix={arXiv},
      primaryClass={astro-ph.IM}
}

@article{Kelly_2004,
	doi = {10.1086/380934},
	url = {https://doi.org/10.1086%2F380934},
	year = 2004,
	month = {feb},
	publisher = {American Astronomical Society},
	volume = {127},
	number = {2},
	pages = {625--645},
	author = {Brandon C. Kelly and Timothy A. McKay},
	title = {Morphological Classification of Galaxies by Shapelet Decomposition in the Sloan Digital Sky Survey},
	journal = {The Astronomical Journal}
}

@ARTICLE{Ramaprakash:2019,
       author = {{Ramaprakash}, A.~N. and {Rajarshi}, C.~V. and {Das}, H.~K. and {Khodade}, P. and {Modi}, D. and {Panopoulou}, G. and {Maharana}, S. and {Blinov}, D. and {Angelakis}, E. and {Casadio}, C. and {Fuhrmann}, L. and {Hovatta}, T. and {Kiehlmann}, S. and {King}, O.~G. and {Kylafis}, N. and {Kougentakis}, A. and {Kus}, A. and {Mahabal}, A. and {Marecki}, A. and {Myserlis}, I. and {Paterakis}, G. and {Paleologou}, E. and {Liodakis}, I. and {Papadakis}, I. and {Papamastorakis}, I. and {Pavlidou}, V. and {Pazderski}, E. and {Pearson}, T.~J. and {Readhead}, A.~C.~S. and {Reig}, P. and {S{\l}owikowska}, A. and {Tassis}, K. and {Zensus}, J.~A.},
        title = "{RoboPol: a four-channel optical imaging polarimeter}",
      journal = {\mnras},
         year = 2019,
        month = may,
       volume = {485},
       number = {2},
        pages = {2355-2366},
          doi = {10.1093/mnras/stz557},
archivePrefix = {arXiv},
       eprint = {1902.08367},
 primaryClass = {astro-ph.IM},
       adsurl = {https://ui.adsabs.harvard.edu/abs/2019MNRAS.485.2355R}
}

@ARTICLE{Handler:2003,
       author = {{Handler}, G.},
        title = "{Combining Aperture and PSF-Fitting Photometry}",
      journal = {Baltic Astronomy},
         year = 2003,
        month = jan,
       volume = {12},
        pages = {243-246},
          doi = {10.1515/astro-2017-0047},
       adsurl = {https://ui.adsabs.harvard.edu/abs/2003BaltA..12..243H}
}

@ARTICLE{Becker:2007,
       author = {{Becker}, Andrew C. and {Silvestri}, Nicole M. and {Owen}, Russell E. and {Ivezi{\'c}}, {\v{Z}}eljko and {Lupton}, Robert H.},
        title = "{In Pursuit of LSST Science Requirements: A Comparison of Photometry Algorithms}",
      journal = {\pasp},
         year = 2007,
        month = dec,
       volume = {119},
       number = {862},
        pages = {1462-1482},
          doi = {10.1086/524710},
archivePrefix = {arXiv},
       eprint = {0712.0637},
 primaryClass = {astro-ph},
       adsurl = {https://ui.adsabs.harvard.edu/abs/2007PASP..119.1462B}
}

@ARTICLE{Massey:2005,
       author = {{Massey}, Richard and {Refregier}, Alexandre},
        title = "{Polar shapelets}",
      journal = {\mnras},
         year = 2005,
        month = oct,
       volume = {363},
       number = {1},
        pages = {197-210},
          doi = {10.1111/j.1365-2966.2005.09453.x},
archivePrefix = {arXiv},
       eprint = {astro-ph/0408445},
 primaryClass = {astro-ph},
       adsurl = {https://ui.adsabs.harvard.edu/abs/2005MNRAS.363..197M}
}

@ARTICLE{Stetson:1987,
       author = {{Stetson}, Peter B.},
        title = "{DAOPHOT: A Computer Program for Crowded-Field Stellar Photometry}",
      journal = {\pasp},
         year = 1987,
        month = mar,
       volume = {99},
        pages = {191},
          doi = {10.1086/131977},
       adsurl = {https://ui.adsabs.harvard.edu/abs/1987PASP...99..191S}
}

@ARTICLE{Kypriotakis:2024,
       author = {{Kypriotakis}, John Andrew and {Maharana}, Siddharth and {Anche}, Ramya M. and {Rajarshi}, Chaitanya V. and {Ramaprakash}, Anamparambu and {Joshi}, Bhushan and {Basyrov}, Artem and {Blinov}, Dmitry and {Ghosh}, Tuhin and {Gjerl{\o}w}, Eirik and {Kiehlmann}, Sebastian and {Mandarakas}, Nikolaos and {Panopoulou}, Georgia V. and {Papadaki}, Katerina and {Pavlidou}, Vasiliki and {Pearson}, Timothy J. and {Pelgrims}, Vincent and {Potter}, Stephen B. and {Readhead}, Anthony C.~S. and {Skalidis}, Raphael and {Tassis}, Konstantinos},
        title = "{Wide Area Linear Optical Polarimeter North instrument I: optical design, filter design, and calibration}",
      journal = {Journal of Astronomical Telescopes, Instruments, and Systems},
         year = 2024,
        month = oct,
       volume = {10},
          eid = {044005},
        pages = {044005},
          doi = {10.1117/1.JATIS.10.4.044005},
archivePrefix = {arXiv},
       eprint = {2412.00964},
 primaryClass = {astro-ph.IM},
       adsurl = {https://ui.adsabs.harvard.edu/abs/2024JATIS..10d4005K}
}

@ARTICLE{Mighell:2005,
       author = {{Mighell}, Kenneth J.},
        title = "{Stellar photometry and astrometry with discrete point spread functions}",
      journal = {\mnras},
         year = 2005,
        month = aug,
       volume = {361},
       number = {3},
        pages = {861-878},
          doi = {10.1111/j.1365-2966.2005.09208.x},
archivePrefix = {arXiv},
       eprint = {astro-ph/0505455},
 primaryClass = {astro-ph},
       adsurl = {https://ui.adsabs.harvard.edu/abs/2005MNRAS.361..861M}
}

@ARTICLE{Anderson:2000,
       author = {{Anderson}, Jay and {King}, Ivan R.},
        title = "{Toward High-Precision Astrometry with WFPC2. I. Deriving an Accurate Point-Spread Function}",
      journal = {\pasp},
         year = 2000,
        month = oct,
       volume = {112},
       number = {776},
        pages = {1360-1382},
          doi = {10.1086/316632},
archivePrefix = {arXiv},
       eprint = {astro-ph/0006325},
 primaryClass = {astro-ph},
       adsurl = {https://ui.adsabs.harvard.edu/abs/2000PASP..112.1360A}
}

@ARTICLE{2013A&A...551A.119P,
       author = {{Piotrowski}, L.~W. and {Batsch}, T. and {Czyrkowski}, H. and {Cwiok}, M. and {Dabrowski}, R. and {Kasprowicz}, G. and {Majcher}, A. and {Majczyna}, A. and {Malek}, K. and {Mankiewicz}, L. and {Nawrocki}, K. and {Opiela}, R. and {Siudek}, M. and {Sokolowski}, M. and {Wawrzaszek}, R. and {Wrochna}, G. and {Zaremba}, M. and {{\.Z}arnecki}, A.~F.},
        title = "{PSF modelling for very wide-field CCD astronomy}",
      journal = {\aap},
         year = 2013,
        month = mar,
       volume = {551},
          eid = {A119},
        pages = {A119},
          doi = {10.1051/0004-6361/201219230},
archivePrefix = {arXiv},
       eprint = {1302.0145},
 primaryClass = {astro-ph.IM},
       adsurl = {https://ui.adsabs.harvard.edu/abs/2013A&A...551A.119P}
}

@INPROCEEDINGS{Mighell+1999,
       author = {{Mighell}, Kenneth J.},
        title = "{Algorithms for CCD Stellar Photometry}",
    booktitle = {Astronomical Data Analysis Software and Systems VIII},
         year = 1999,
       editor = {{Mehringer}, David M. and {Plante}, Raymond L. and {Roberts}, Douglas A.},
       series = {Astronomical Society of the Pacific Conference Series},
       volume = {172},
        month = jan,
        pages = {317},
       adsurl = {https://ui.adsabs.harvard.edu/abs/1999ASPC..172..317M}
}

@MISC{van2014scikit,
  title={scikit-image: image processing in Python},
  author={Van der Walt, Stefan and Sch{\"o}nberger, Johannes L and Nunez-Iglesias, Juan and Boulogne, Fran{\c{c}}ois and Warner, Joshua D and Yager, Neil and Gouillart, Emmanuelle and Yu, Tony},
  journal={PeerJ},
  volume={2},
  pages={e453},
  year={2014},
  publisher={PeerJ Inc.}
}

@MISC{larry_bradley_2020_4044744,
      title = "{astropy/photutils: 1.0.0}",
      author = {{Bradley}, Larry and {Sip{\H{o}}cz}, Brigitta and {Robitaille}, Thomas and {Tollerud}, Erik and {Vin{\'\i}cius}, Z{\'e} and {Deil}, Christoph and {Barbary}, Kyle and {Wilson}, Tom J and {Busko}, Ivo and {G{\"u}nther}, Hans Moritz and {Cara}, Mihai and {Conseil}, Simon and {Bostroem}, Azalee and {Droettboom}, Michael and {Bray}, E.~M. and {Andersen Bratholm}, Lars and {Lim}, P.~L. and {Barentsen}, Geert and {Craig}, Matt and {Pascual}, Sergio and {Perren}, Gabriel and {Greco}, Johnny and {Donath}, Axel and {De Val-Borro}, Miguel and {Kerzendorf}, Wolfgang and {Bach}, Yoonsoo P. and {Weaver}, Benjamin Alan and {D'Eugenio}, Francesco and {Souchereau}, Harrison and {Ferreira}, Leonardo},
         journal = {Zenodo},
         year = 2020,
        month = sep,
          eid = {10.5281/zenodo.4044744},
          doi = {10.5281/zenodo.4044744},
      version = {1.0.0},
    publisher = {Zenodo},
       adsurl = {https://ui.adsabs.harvard.edu/abs/2020zndo...4044744B}
}

@INPROCEEDINGS{2001ASPC..238..269L,
       author = {{Lupton}, R. and {Gunn}, J.~E. and {Ivezi{\'c}}, Z. and {Knapp}, G.~R. and {Kent}, S.},
        title = "{The SDSS Imaging Pipelines}",
    booktitle = {Astronomical Data Analysis Software and Systems X},
         year = 2001,
       editor = {{Harnden}, F.~R., Jr. and {Primini}, Frances A. and {Payne}, Harry E.},
       series = {Astronomical Society of the Pacific Conference Series},
       volume = {238},
        month = jan,
        pages = {269},
archivePrefix = {arXiv},
       eprint = {astro-ph/0101420},
 primaryClass = {astro-ph},
       adsurl = {https://ui.adsabs.harvard.edu/abs/2001ASPC..238..269L}
}

@ARTICLE{2003MNRAS.338...35R,
       author = {{Refregier}, Alexandre},
        title = "{Shapelets - I. A method for image analysis}",
      journal = {\mnras},
         year = 2003,
        month = jan,
       volume = {338},
       number = {1},
        pages = {35-47},
          doi = {10.1046/j.1365-8711.2003.05901.x},
archivePrefix = {arXiv},
       eprint = {astro-ph/0105178},
 primaryClass = {astro-ph},
       adsurl = {https://ui.adsabs.harvard.edu/abs/2003MNRAS.338...35R}
}

@ARTICLE{2003MNRAS.338...48R,
       author = {{Refregier}, Alexandre and {Bacon}, David},
        title = "{Shapelets - II. A method for weak lensing measurements}",
      journal = {\mnras},
         year = 2003,
        month = jan,
       volume = {338},
       number = {1},
        pages = {48-56},
          doi = {10.1046/j.1365-8711.2003.05902.x},
archivePrefix = {arXiv},
       eprint = {astro-ph/0105179},
 primaryClass = {astro-ph},
       adsurl = {https://ui.adsabs.harvard.edu/abs/2003MNRAS.338...48R}
}

@ARTICLE{Maharana:2022,
       author = {{Maharana}, Siddharth and {Anche}, Ramya M. and {Ramaprakash}, Anamparambu N. and {Joshi}, Bhushan and {Basyrov}, Artem and {Blinov}, Dmitry and {Casadio}, Carolina and {Deka}, Kishan and {Eriksen}, Hans Kristian and {Ghosh}, Tuhin and {Gjerl{\o}w}, Eirik and {Kypriotakis}, John A. and {Kiehlmann}, Sebastian and {Mandarakas}, Nikolaos and {Panopoulou}, Georgia V. and {Papadaki}, Katerina and {Pavlidou}, Vasiliki and {Pearson}, Timothy J. and {Pelgrims}, Vincent and {Potter}, Stephen B. and {Readhead}, Anthony C.~S. and {Skalidis}, Raphael and {Svalheim}, Trygve Leithe and {Tassis}, Konstantinos and {Wehus}, Ingunn K.},
        title = "{WALOP-South: a four-camera one-shot imaging polarimeter for PASIPHAE survey. Paper II - polarimetric modeling and calibration}",
      journal = {Journal of Astronomical Telescopes, Instruments, and Systems},
         year = 2022,
        month = jul,
       volume = {8},
          eid = {038004},
        pages = {038004},
          doi = {10.1117/1.JATIS.8.3.038004},
archivePrefix = {arXiv},
       eprint = {2208.12441},
 primaryClass = {astro-ph.IM},
       adsurl = {https://ui.adsabs.harvard.edu/abs/2022JATIS...8c8004M}
}

@ARTICLE{Maharana:2024,
       author = {{Maharana}, Siddharth and {Ramaprakash}, A.~N. and {Rajarshi}, Chaitanya and {Khodade}, Pravin and {Joshi}, Bhushan and {Chordia}, Pravin and {Kohok}, Abhay and {Anche}, Ramya M. and {Modi}, Deepa and {Kypriotakis}, John A. and {Deokar}, Amit and {Kinjawadekar}, Aditya and {Potter}, Stephen B. and {Blinov}, Dmitry and {Eriksen}, Hans Kristian and {Falalaki}, Myrto and {Gajjar}, Hitesh and {Ghosh}, Tuhin and {Gjerl{\o}w}, Eirik and {Kiehlmann}, Sebastain and {Liodakis}, Ioannis and {Mandarakas}, Nikolaos and {Panopoulou}, Georgia V. and {Pavlidou}, Vasiliki and {Pearson}, Timothy J. and {Pelgrims}, Vincent and {Readhead}, Anthony C.~S. and {Skalidis}, Raphael and {Tassis}, Konstantinos and {Uppal}, Namita and {Wehus}, Ingunn K.},
        title = "{Systems design, assembly, integration and lab testing of WALOP-South Polarimeter}",
      journal = {arXiv e-prints},
         year = 2024,
        month = jun,
          eid = {arXiv:2406.19428},
        pages = {arXiv:2406.19428},
          doi = {10.48550/arXiv.2406.19428},
archivePrefix = {arXiv},
       eprint = {2406.19428},
 primaryClass = {astro-ph.IM},
       adsurl = {https://ui.adsabs.harvard.edu/abs/2024arXiv240619428M}
}

@ARTICLE{Maharana:2021,
       author = {{Maharana}, Siddharth and {Kypriotakis}, John A. and {Ramaprakash}, Anamparambu N. and {Rajarshi}, Chaitanya and {Anche}, Ramya M. and {Shrish} and {Blinov}, Dmitry and {Eriksen}, Hans Kristian and {Ghosh}, Tuhin and {Gjerl{\o}w}, Eirik and {Mandarakas}, Nikolaos and {Panopoulou}, Georgia V. and {Pavlidou}, Vasiliki and {Pearson}, Timothy J. and {Pelgrims}, Vincent and {Potter}, Stephen B. and {Readhead}, Anthony C.~S. and {Skalidis}, Raphael and {Tassis}, Konstantinos and {Wehus}, Ingunn K.},
        title = "{WALOP-South: a four-camera one-shot imaging polarimeter for PASIPHAE survey. Paper I{\textemdash}optical design}",
      journal = {Journal of Astronomical Telescopes, Instruments, and Systems},
         year = 2021,
        month = jan,
       volume = {7},
          eid = {014004},
        pages = {014004},
          doi = {10.1117/1.JATIS.7.1.014004},
archivePrefix = {arXiv},
       eprint = {2102.09505},
 primaryClass = {astro-ph.IM},
       adsurl = {https://ui.adsabs.harvard.edu/abs/2021JATIS...7a4004M}
}

@inproceedings{walop_s_spie_2020,
author = {Siddharth Maharana and John A. Kypriotakis and A. N. Ramaprakash and Pravin Khodade and Chaitanya Rajarshi and Bhushan S. Joshi and Pravin Chordia and Ramya M. Anche and Shrish Mishra and Dmitry Blinov and Hans Kristian Eriksen and Tuhin Ghosh and Eirik Gjerløw and Nikolaos Mandarakas and Georgia V. Panopoulou and Vasiliki Pavlidou and Timothy J. Pearson and Vincent Pelgrims and Stephen B. Potter and Anthony C. S. Readhead and Raphail Skalidis and Konstantinos Tassis and Ingunn K. Wehus},
title = {{WALOP-South: A wide-field one-shot linear optical polarimeter for PASIPHAE survey}},
volume = {11447},
booktitle = {Ground-based and Airborne Instrumentation for Astronomy VIII},
editor = {Christopher J. Evans and Julia J. Bryant and Kentaro Motohara},
organization = {International Society for Optics and Photonics},
publisher = {SPIE},
pages = {1135 -- 1146},
year = {2020},
doi = {10.1117/12.2554396},
URL = {https://doi.org/10.1117/12.2554396}
}

@book{Bracewell2000,
  author    = {Bracewell, R. N.},
  title     = {The Fourier Transform and Its Applications},
  publisher = {McGraw-Hill},
  year      = {2000}
}

@article{Guizar-Sicairos:08,
author = {Manuel Guizar-Sicairos and Samuel T. Thurman and James R. Fienup},
journal = {Opt. Lett.},
number = {2},
pages = {156--158},
publisher = {Optica Publishing Group},
title = {Efficient subpixel image registration algorithms},
volume = {33},
month = {Jan},
year = {2008},
url = {https://opg.optica.org/ol/abstract.cfm?URI=ol-33-2-156},
doi = {10.1364/OL.33.000156},
}

@article{Powell1964,
    author = {Powell, M. J. D.},
    title = {An efficient method for finding the minimum of a function of several variables without calculating derivatives},
    journal = {The Computer Journal},
    volume = {7},
    number = {2},
    pages = {155-162},
    year = {1964},
    month = {01},
    doi = {10.1093/comjnl/7.2.155},
    url = {https://doi.org/10.1093/comjnl/7.2.155},
    eprint = {https://academic.oup.com/comjnl/article-pdf/7/2/155/959784/070155.pdf},
}

@article{Virtanen2020,
  author  = {Virtanen, P. et al.},
  title   = {{SciPy} 1.0: Fundamental Algorithms for Scientific Computing in Python},
  journal = {Nature Methods},
  volume  = {17},
  pages   = {261--272},
  year    = {2020},
  doi     = {10.1038/s41592-019-0686-2}
}

@article{Gordon_2008,
       author = {{Gordon}, Karl D. and {Engelbracht}, Charles W. and {Rieke}, George H. and {Misselt}, K.~A. and {Smith}, J. -D.~T. and {Kennicutt}, Robert C., Jr.},
        title = "{The Behavior of the Aromatic Features in M101 H II Regions: Evidence for Dust Processing}",
      journal = {\apj},
         year = 2008,
        month = jul,
       volume = {682},
       number = {1},
        pages = {336-354},
          doi = {10.1086/589567},
archivePrefix = {arXiv},
       eprint = {0804.3223},
 primaryClass = {astro-ph},
       adsurl = {https://ui.adsabs.harvard.edu/abs/2008ApJ...682..336G}
      }

@ARTICLE{Aniano_2011,
       author = {{Aniano}, G. and {Draine}, B.~T. and {Gordon}, K.~D. and {Sandstrom}, K.},
        title = "{Common-Resolution Convolution Kernels for Space- and Ground-Based Telescopes}",
      journal = {\pasp},
         year = 2011,
        month = oct,
       volume = {123},
       number = {908},
        pages = {1218},
          doi = {10.1086/662219},
archivePrefix = {arXiv},
       eprint = {1106.5065},
 primaryClass = {astro-ph.IM},
       adsurl = {https://ui.adsabs.harvard.edu/abs/2011PASP..123.1218A}
}

@ARTICLE{Richardson_1972,
       author = {{Richardson}, William Hadley},
        title = "{Bayesian-Based Iterative Method of Image Restoration}",
      journal = {Journal of the Optical Society of America (1917-1983)},
         year = 1972,
        month = jan,
       volume = {62},
       number = {1},
        pages = {55},
          doi = {10.1364/JOSA.62.000055},
       adsurl = {https://ui.adsabs.harvard.edu/abs/1972JOSA...62...55R}
}

@ARTICLE{Lucy_1974,
       author = {{Lucy}, L.~B.},
        title = "{An iterative technique for the rectification of observed distributions}",
      journal = {\aj},
         year = 1974,
        month = jun,
       volume = {79},
        pages = {745},
          doi = {10.1086/111605},
       adsurl = {https://ui.adsabs.harvard.edu/abs/1974AJ.....79..745L}
}

@article{Wu_2025,
doi = {10.1088/1674-4527/ae088a},
url = {https://doi.org/10.1088/1674-4527/ae088a},
year = {2025},
month = {oct},
publisher = {National Astromonical Observatories, CAS and IOP Publishing},
volume = {25},
number = {12},
pages = {125003},
author = {Wu, You and Li, Nan and Shan, Huan-Yuan and Wei, Peng and Wei, Cheng-Liang and Nie, Lin and Ren, Juan-Juan and Ban, Zhang and Li, Xiao-Bo and Yang, Xun and Jiang, Yu-Xi and Ma, Hong-Cai and Wang, Wei and Liu, Chao},
title = {Improving PSF Reconstruction for CSST: A Combined Approach with Deep Learning Source Selection and Empirical Correction},
journal = {Research in Astronomy and Astrophysics}
}

@ARTICLE{Lauer_1999,
       author = {{Lauer}, Tod R.},
        title = "{The Photometry of Undersampled Point-Spread Functions}",
      journal = {\pasp},
         year = 1999,
        month = nov,
       volume = {111},
       number = {765},
        pages = {1434-1443},
          doi = {10.1086/316460},
archivePrefix = {arXiv},
       eprint = {astro-ph/9907100},
 primaryClass = {astro-ph},
       adsurl = {https://ui.adsabs.harvard.edu/abs/1999PASP..111.1434L}
}

@ARTICLE{Barron_2007,
       author = {{Barron}, N. and {Borysow}, M. and {Beyerlein}, K. and {Brown}, M. and {Lorenzon}, W. and {Schubnell}, M. and {Tarl{\'e}}, G. and {Tomasch}, A. and {Weaverdyck}, C.},
        title = "{Subpixel Response Measurement of Near-Infrared Detectors}",
      journal = {\pasp},
         year = 2007,
        month = apr,
       volume = {119},
       number = {854},
        pages = {466-475},
          doi = {10.1086/517620},
archivePrefix = {arXiv},
       eprint = {astro-ph/0611339},
 primaryClass = {astro-ph},
       adsurl = {https://ui.adsabs.harvard.edu/abs/2007PASP..119..466B}
}

@ARTICLE{SExtractor+1996,
       author = {{Bertin}, E. and {Arnouts}, S.},
        title = "{SExtractor: Software for source extraction.}",
      journal = {\aaps},
         year = 1996,
        month = jun,
       volume = {117},
        pages = {393-404},
          doi = {10.1051/aas:1996164},
       adsurl = {https://ui.adsabs.harvard.edu/abs/1996A&AS..117..393B}
}

@INPROCEEDINGS{heasley1999point,
       author = {{Heasley}, J.~N.},
        title = "{Point-Spread Function Fitting Photometry}",
    booktitle = {Precision CCD Photometry},
         year = 1999,
       editor = {{Craine}, Eric R. and {Crawford}, David L. and {Tucker}, Roy A.},
       series = {Astronomical Society of the Pacific Conference Series},
       volume = {189},
        month = jan,
        pages = {56},
       adsurl = {https://ui.adsabs.harvard.edu/abs/1999ASPC..189...56H}
}

\appendix

\section{Polar Shapelet basis}\label{appendix:A}

Figure~\ref{fig:psbasis} shows the real and imaginary parts of the polar shapelet 
basis functions up to order $n_{\rm max} = 5$, together with the central terms 
for $n = 6$ to $n = 34$.

\begin{figure}[!htbp]
\centering
\includegraphics[clip, trim=.1cm .5cm .1cm 1cm, width=\linewidth]{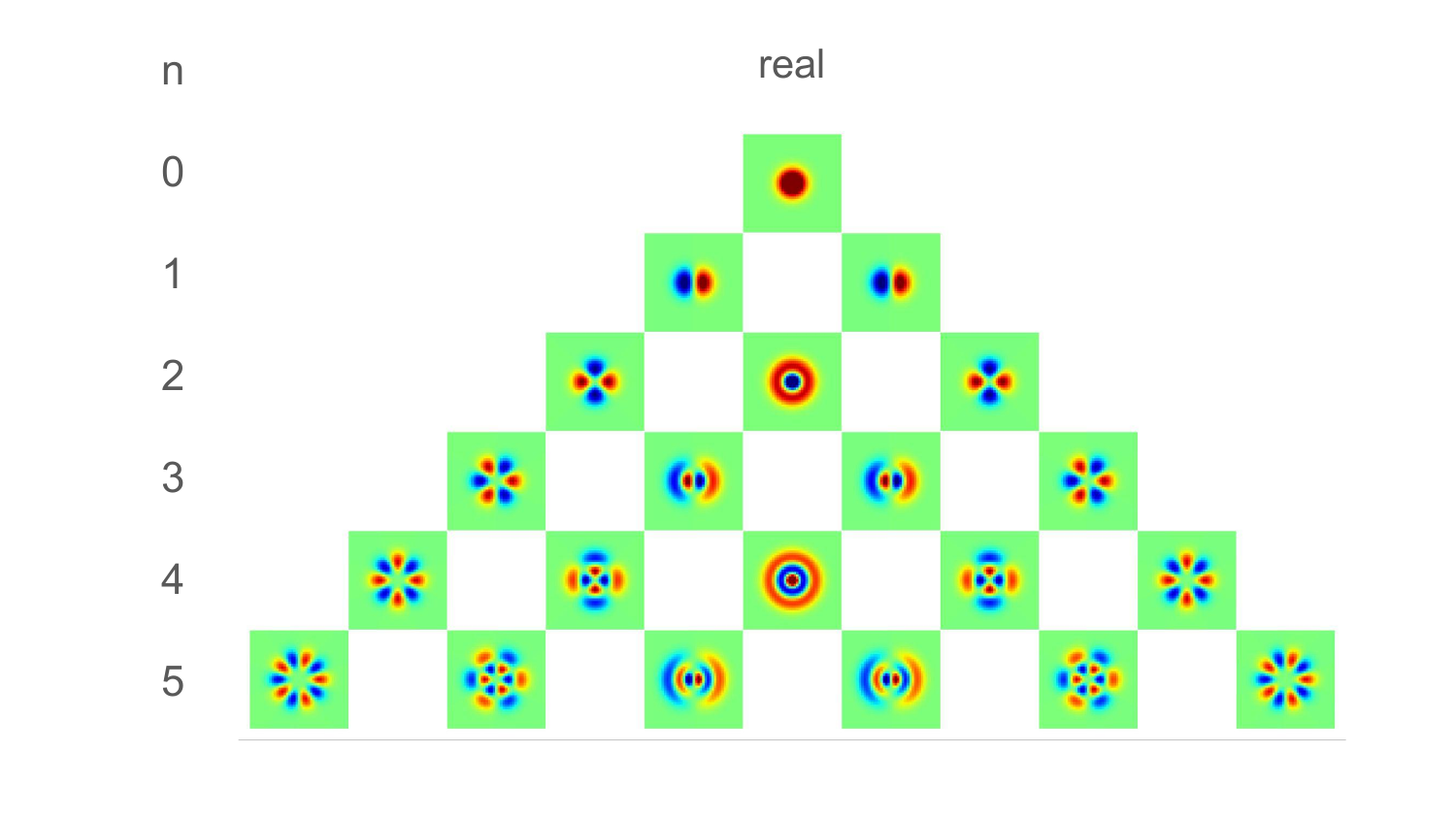}
\includegraphics[clip, trim=.1cm .05cm .1cm 1cm, width=\linewidth]{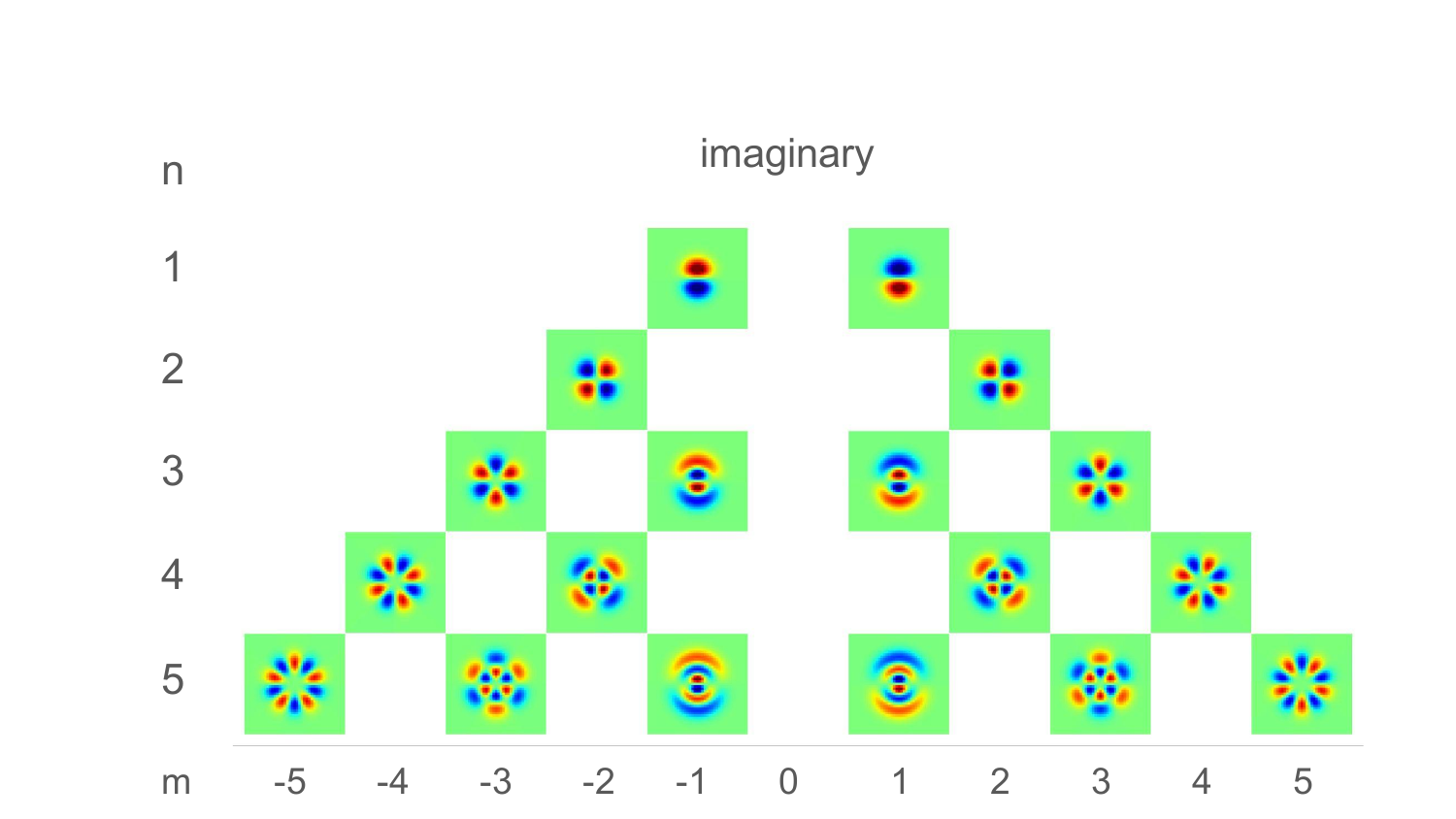}
%\vspace{-30pt}
\includegraphics[width=0.95\hsize]{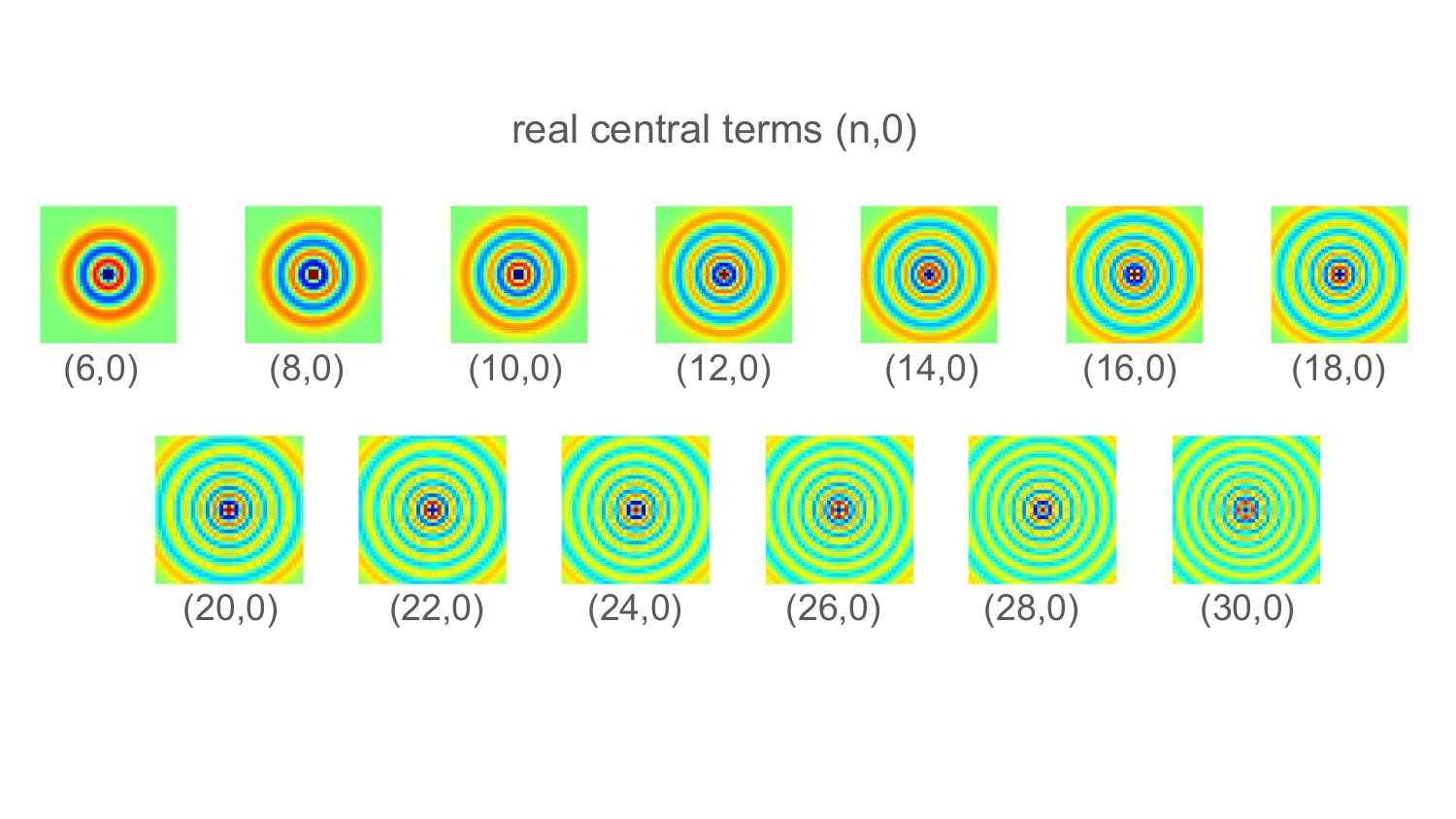}
\vspace{-30pt}
\caption{Real and imaginary components of the shapelet basis functions for various values of $m$ and $n$.}
\label{fig:psbasis}
\end{figure}

\section{PSF photometry on RoboPol data}\label{appendix:B}

The PSF photometry method is applied to starlight observations from the RoboPol survey \citep{Ramaprakash:2019} to assess its performance and compare it with aperture photometry. The RoboPol instrument is a four-channel imaging polarimeter that produces four images of each source in the field-of-view on a $2048 \times 2048$ pixel CCD. The target star is placed at the centre, and a mask is used to suppress background contamination in the central region. A $200 \times 200$ pixel cut-out containing all four images (or spots) of the target star is extracted from the central mask region; the spots are numbered 1 to 4 as shown in Fig.~\ref{fig:cut_out_robopol}. For the analysis, observations of the star PG\_1633+099B obtained on 19 May 2017 with an exposure time of $40\,\mathrm{s}$ are used. Each spot is further extracted as an individual star image of $41 \times 41$ pixels, centred on the brightest pixel. Since the seeing conditions differ from those of the WALOP-South simulated images, the $\beta$ parameter is first determined for each spot individually. For the PSF modelling, the mean $\beta= 2.2$ in pixel value across the four spots is adopted, and the analysis is performed using 33  shapelet coefficients in total. A model PSF is constructed from the polar shapelet basis functions for each of the four spots, and the respective PSF model is applied to each spot to extract the total photon count. Aperture photometry is performed using a circular aperture of radius $r = 12$ pixels, which encloses more than $99\%$ of the total photon count. Both methods yield consistent values of the total photon count, but a reduced photometric uncertainty is obtained with PSF photometry. The results from both methods are listed in \cref{tab:phot_compare}. The mean photometric accuracy achieved with PSF photometry is ${\sim}0.35\%$. Fig.~\ref{fig:robopol_psf} shows the observed image, the best-fit model, and the residuals for the RoboPol data. The model accurately recovers the total photon counts, with no structured residuals in the fit.

\begin{figure}
\centering
\includegraphics[width=1.08\hsize]{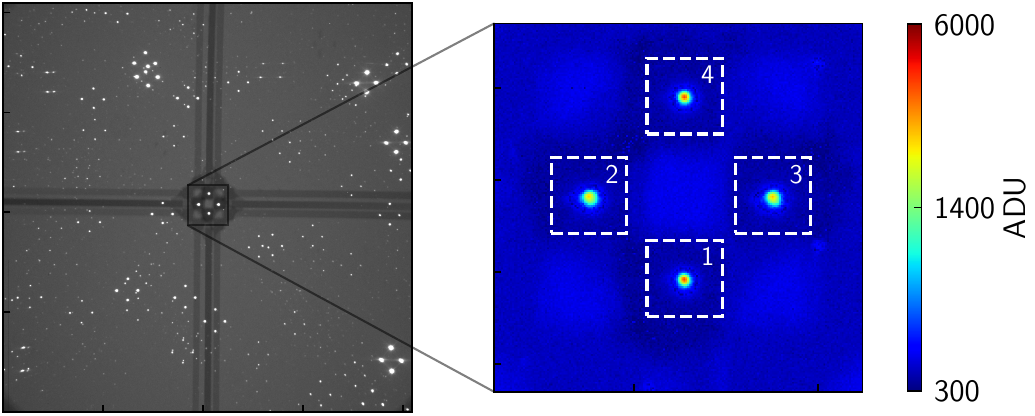}
\caption{(\textit{Left panel}) A RoboPol image of the star PG\_1633+099B. Each source produces four images on the CCD. The target star is placed at the centre, and a mask supported by four arms is used to suppress background contamination in the central region. (\textit{Right panel}) A zoomed-in cut-out of the central mask region.} 
\label{fig:cut_out_robopol}
\end{figure}

\begin{figure}
\centering
\includegraphics[width=\hsize]{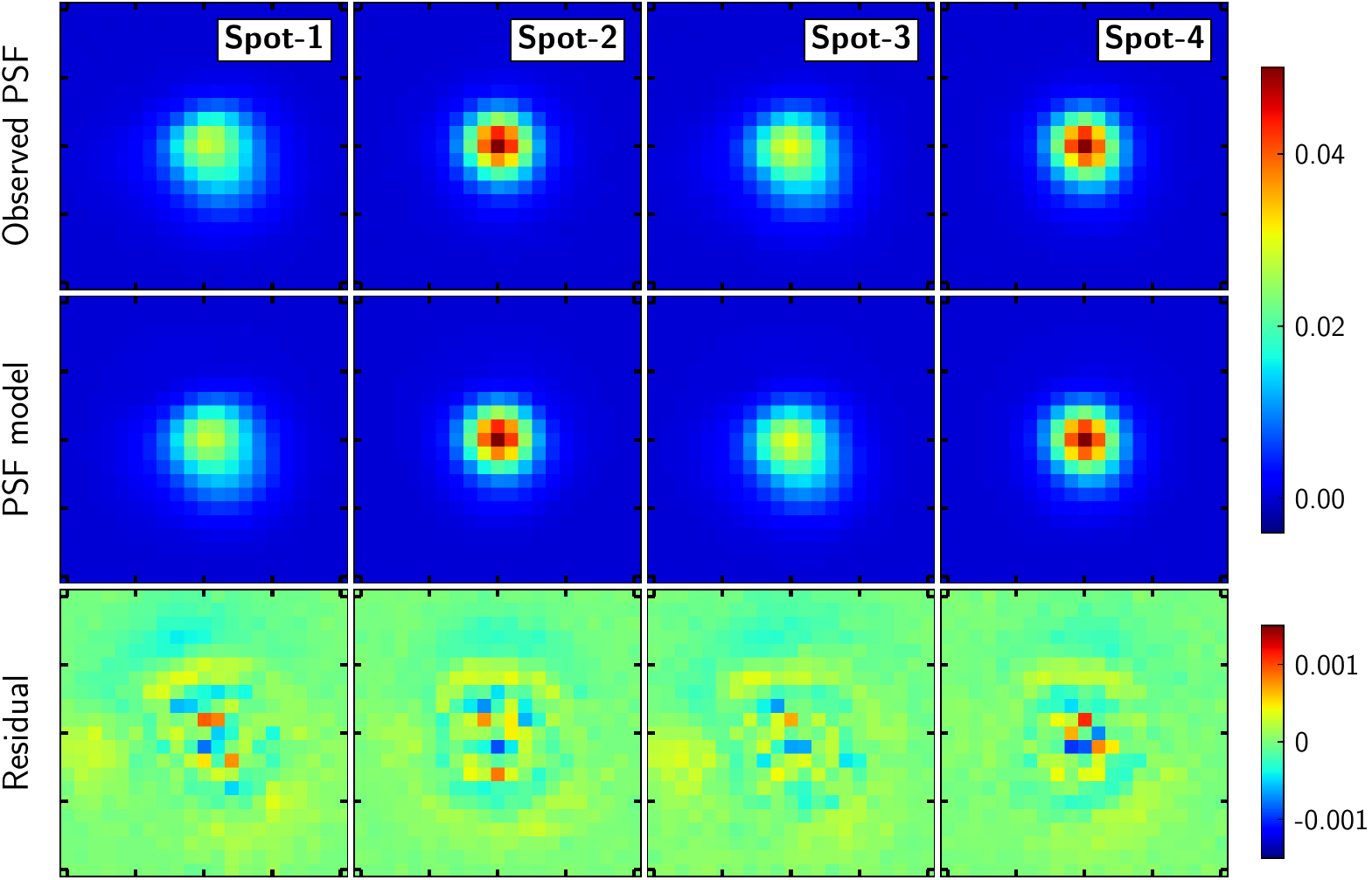}
\caption{RoboPol observation of the star PG\_1633+099B. \textit{Top row}: background-subtracted star images; \textit{middle row}: best-fit model constructed using polar shapelets; \textit{bottom row}: fit residuals.}
\label{fig:robopol_psf}
\end{figure}

\begin{table}
	\centering
	\caption{Comparison of performance between PSF photometry and aperture photometry.}
	\begin{tabular}{ccc}
	\hline \hline
	            &   PSF photometry counts & Aperture photometry counts  \\
	            &  		(in ADU) & (in ADU)  \\ 
	\hline
	Spot-1 & $146060 \pm 510$ & $146161 \pm 697$ \\
	Spot-2 & $142176 \pm 494$ & $141701 \pm 651$ \\
	Spot-3 & $146618 \pm 511$ & $145909 \pm 697$ \\
	Spot-4 & $141799 \pm 493$ & $141208 \pm 623$ \\
	\hline
\end{tabular}
\label{tab:phot_compare}
\end{table}

\label{lastpage}
\end{document}